\documentstyle[aps,epsfig]{revtex}

\begin{document}

\title{Subthreshold $\rho^0$ photoproduction on ${^3}He$}

\author{M.A. Kagarlis\cite{kag}, Z. Papandreou, G.M. Huber, G.J. Lolos, A.
Shinozaki, E.J. Brash, F. Farzanpay, M. Iurescu, and A.~Weinerman}
\address{Department of Physics, University of Regina, Regina, Saskatchewan,
S4S 0A2, Canada}

\author{G. Garino\cite{garino}, and K. Maruyama}
\address{Institute for Nuclear Study\cite{ins}, University of Tokyo, Tanashi,
Tokyo 188, Japan}

\author{O. Konno, K. Maeda, T. Terasawa, and H. Yamazaki}
\address{Department of Physics, Tohoku University, Sendai 980, Japan}

\author{T. Emura, H. Hirosawa, K. Niwa, and H. Yamashita}
\address{Department of Applied Physics, Tokyo University of Agriculture and
Technology, Koganei, Tokyo 184, Japan}

\author{S. Endo, K. Miyamoto, and Y. Sumi}
\address{Department of Physics, Hiroshima University, Higashi-Hiroshima 724,
Japan}

\author{A. Leone, and R. Perrino}
\address{INFN-Sezione di Lecce, I-73100 Lecce, Italy}

\author{T. Maki }
\address{University of Occupational and Environmental Health, Kitakyushu 807,
Japan}

\author{A. Sasaki}
\address{College of General Education, Akita University, Akita 010, Japan}

\author{J.C. Kim}
\address{Department of Physics, Seoul National University, Seoul 151-742,
Korea}

\author{(The TAGX Collaboration)}

\date{\today}
\maketitle

\begin{abstract}
A large reduction of the $\rho^0$ mass in the nuclear medium is
reported, inferred from dipion photoproduction spectra in the 1 GeV
region, for the reaction ${^3}He$($\gamma$,$\pi^+\pi^-$)X with a 10\%
duty factor tagged-photon beam and the TAGX multi-particle
spectrometer. The energy range covered (800$\le$E$_\gamma$$\le$1120
MeV) lies mostly below the free $\rho^0$ production threshold, a
region which is believed sensitive to modifications of light
vector-meson properties at nuclear-matter densities.  The $\rho^0$
masses extracted from the MC fitting of the data, m$^*_{\rho^0}$ =
642$\pm$40, 669$\pm$32, and 682$\pm$56 MeV/c$^2$ for E$_\gamma$ in the
800-880, 880-960, and 960-1040 MeV regions respectively, are
independently corroborated by a measured, assumption-free, kinematical
observable. This mass shift, far exceeding current mean-field driven
theoretical predictions, may be suggestive of $\rho^0$ decay within
the range of the nucleonic field.
\end{abstract}

\pacs{13.60.Le, 25.20.Lj, 12.38.Qk, 14.40.Cs}

\section{Introduction}
\label{sec:intro}

Hadronic dynamics have until recently been comprising of two
non-overlapping domains, distinctly separated along the lines of their
respective description of matter as hadronic or quark. On the
low-energy side of the spectrum, matter is probed on a scale where
pions and baryonic resonances are the relevant constituents. These
have been rather succesfully employed by quasi-phenomenological models
in providing prescriptions for free and in-medium hadronic
interactions. At the higher-energy end, far above energies typical of
baryonic resonances, quark degrees of freedom become accessible and
matter tends asymptotically towards quark-gluon plasma. In this
regime, described by perturbative QCD, quarks become deconfined and
chiral symmetry is restored.

Although the origin of quark confinement is not known, it is evident
that, whatever the reasons, it must induce spontaneous breaking of
chiral symmetry \cite{casher}. The two phenomena are therefore linked
and in the limit of chiral symmetry restoration, as quark masses tend
to zero, vector-meson masses and widths are also expected to change
\cite{pisar}.  The transition from hadronic to quark matter and the
effect of increasing density and temperature on the properties of
light vector mesons have been addressed in the context of QCD sum
rules \cite{hat} as well as effective Lagrangians \cite{geb1}. In
particular, a temperature and density dependent lowering of the
$\rho^0$ mass is regarded as a precursor of the chiral phase
transition, expected to be measurable even at normal nuclear density
\cite{geb2}. Thus, the energy region from hadronic phenomenology to
the domain of perturbative QCD has come to the foreground, as the
study of vector-meson property modifications in this range appears to
be holding the key to understanding the mechanism of chiral symmetry
restoration.

The recent advent of high duty-cycle photon beams, on one hand, and
high-resolution dilepton spectrometers, on the other, have opened up
the possibility of reconciliation between the hadronic and quark
depictions of matter. Hadronic probes coupled with hadron
spectrometers had been extensively used in the past for energies up to
the $\Delta$ resonance, but higher-up, where multi-pion production
channels increasingly dominate, the combination becomes rapidly
cumbersome due to medium distortions from initial and final state
interactions. For the study of the vector mesons that couple to
multi-pion states, the photon is ideally suited. The recognition of
the electromagnetic interaction as an indispensable probe at either
the entrance or the exit channel, if the vector-meson field in the
nuclear medium is to be delineated, has generated activity in a
variety of fields.

\subsection{Relativistic Heavy Ion Results}

At relativistic energies, under extreme conditions of temperature and
density, the quest for signs of a phase transition from hadronic
matter to quark-gluon plasma is at the heart of experimental programs
using heavy-ion beams (e.g. at CERN, GSI, and RHIC) and dilepton
spectrometers. A series of pioneering studies from the CERES,
HELIOS-3, and NA50 colaborations at SPS/CERN with central S + Au, S +
W, Pb + Au, and Pb + Pb collisions, have largely been interpreted as
indicative of a downward shift of the $\rho^0$ mass in the nuclear
medium \cite{ceres,he3}. The invariant mass spectra for dilepton
production that have been measured in these experiments, when compared
with the respective p-A spectra, show a large enhancement at low
invariant mass regions. The excess dileptons are thought to originate
from the decay of mesonic resonances produced in $\pi^+\pi^-$
annihilation and, in the mass region 0.2$<$m$_{l^+l^-}<$0.5 GeV/c$^2$,
the models advocating vector-meson medium modifications attribute the
enhancement to the decay of the $\rho^0$ meson with a downward-shifted
mass \cite{geb3,cas}.  Nonetheless, an explanation in terms of
conventional phenomenological $\rho^0$-meson medium modifications is
also consistent with the CERN data \cite{chanfray,ccf,friman}.  In
this more conservative approach, the $\rho^0$ spectral function below
0.6 GeV/c$^2$ in dilepton invariant mass is appreciably enhanced from
the contributions of ``rhosobar''-type excitations such as the $\Delta
N^{-1}$, N$^*$(1720)N$^{-1}$ and $\Delta^*(1905) N^{-1}$, a
consequence of the strong coupling of the $\rho^0$ meson with
$\pi^+\pi^-$ states in the nuclear medium.  This being the case, the
downward shift of the $\rho^0$-meson mass in the nuclear medium may
amount to no more than a convenient parametrization
\cite{chanfray}.

From a theoretical perspective, a model combining chiral SU(3)
dynamics with vector-meson dominance in an effective Lagrangian has
shown that chiral restoration does not demand a drastic reduction of
vector-meson masses in the nuclear medium \cite{weise}. The latter
model, in qualitative agreement with the hadronic-phenomenological
models \cite{chanfray,friman}, predicts a substantial enhancement of
the $\rho^0$ spectral density below the nominal resonance mass, with
only a marginal mass reduction. This result is tantamount to the
$\rho^0$ dissolving in the medium. Moreover, both an interpretation of
the CERES data as the outcome of either medium-modified hadronic
interactions, or interactions of dissociated quarks in the quark-gluon
phase, yield remarkably consistent results in the framework of
Ref.~\cite{weise}, leading to the conjecture that both the hadronic
and quark-gluon phases must be present \cite{weise}. The latter
conclusion is also drawn by a synthesis of the mass-scaling
\cite{geb2} and hadronic-phenomenological \cite{friman} models,
leading to the prediction that dynamical-hadronic effects that are
dominant up to about nuclear density, mainly via the highly-collective
N$^*$(1520)N$^{-1}$ state, gradually give way to $\rho^0$-meson mass
scaling as the quark degrees of freedom become increasingly
relevant. In the intermediate crossover region of ``duality'', both
the hadronic and quark-gluon phases of matter are expected to coexist
\cite{rwbr}. Thus, although a consensus has been reached on the issue
of coexistence of the hadronic and quark-gluon phases in the
transition region roughly placed upwards from $\sim$1~GeV, the
question remains as to the extent to which the nuclear medium induces
modifications to the $\rho^0$ mass and width.

Other less directly related experimental conjectures of medium
modifications of the $\rho^0$ mass have been deduced from an anomalous
J/$\psi$ suppression, reported by the NA50 collaboration for Pb + Pb
collisions \cite{jpsi}, enhanced K$^+$-$^{12}C$ scattering cross
sections \cite{kaons}, and an IUCF experiment of polarized proton
scattering on $^{28}$Si, though in the latter the medium
renormalization of the $\rho^0$ mass required for agreement with the
data is inconsistent for different observables measured in the same
experiment \cite{si}.

In summary, the experimental results discussed so far are inconclusive
with regard to the magnitude of the $\rho^0$-meson medium mass
modification, and underline the limitations encountered in complex
processes, where the probe interacts with nuclear matter via the
strong interaction and the channel being investigated may not be
disentangled from conventional medium effects.

\subsection{Electromagnetic Probe Results}

These difficulties are largely overcome with the use of photons which
do not suffer from initial state interactions. The availability of
high-flux photon beams, which have compensated for the low interaction
cross sections with nuclear matter, complemented by wide-angle
multi-particle spectrometers, have made possible a new generation of
experiments. In the E$_\gamma\sim$1 GeV region, matter is probed at
short distances $\le$1 fm, which is at the gateway of the energy range
where vector-meson properties are expected to undergo
modifications. In this domain, baryons and mesons are still the
relevant constituents for the description of matter, yet their quark
content becomes increasingly manifest, a fact which is reflected in
QCD-inspired phenomenological models \cite{isg,sai}. The $\rho^0$
meson is the best candidate among the light vector mesons
$\rho^{0,\pm}$, $\omega$, and $\varphi$ as a probe of medium
modifications, since, due to its short lifetime and decay length (1.3
fm), a large portion of $\rho^0$ mesons produced on nuclei will decay
in the medium.

For effective measurements with the photon as probe in the 1 GeV
region, kinematically complete experiments are required, which in turn
necessitate the use of high-duty-cycle tagged photon beams and
large-acceptance multi-particle spectrometers. These requirements were
met by the TAGX detector \cite{maru} at the Institute for Nuclear
Study (INS), where the TAGX collaboration has completed a series of
experiments with the ${^3}He$($\gamma$,$\pi^+\pi^-$)X reaction and a
10\% duty-cycle tagged photon beam. First, a lower-energy experiment
(380$\le$E$_\gamma\le$700 MeV) measured the single- and
double-$\Delta$ contributions to $\pi^+\pi^-$ production
\cite{watts}. Having established these important non-$\rho^0$ dipion
processes, the kinematics of the ${^3}He$($\gamma$,$\pi^+\pi^-$)X
reaction were investigated in the range 800$\le$E$_\gamma\le$1120 MeV,
aiming at the $\rho^0 \rightarrow \pi^+\pi^-$ channel.

Both the coherent and quasifree $\rho^0$ photoproduction mechanisms
are relevant in the energy region of interest. For a ${^3}He$ target,
the energy threshold for the former is E$_\gamma \approx$ 873 MeV,
whereas the 1.083~GeV energy threshold of the elementary $\rho^0$
photoproduction reaction on a nucleon is lowered for the quasifree
channel in the nuclear medium, as the Fermi momentum of the struck
nucleon may be utilized to bring the $\rho^0$ meson on shell. Coherent
photoproduction on nuclei is characterized by small four-momentum
transfers, resulting in $\rho^0$ mesons which mostly decay outside of
the nucleus \cite{crho}. The latter is, consequently, of limited
utility in probing vector-meson medium modifications.

In contrast, quasifree subthreshold photoproduction (E$_\gamma <$
1.083~GeV) on one hand warrants that the interaction took place in the
interior of the nucleus, since the nucleon Fermi momentum is required
to produce the $\rho^0$, and on the other produces slower $\rho^0$
mesons, more likely to also decay inside the nucleus. Moreover,
subthreshold photoproduction, whether the target nucleus remains bound
in the final state (exclusive channel) or decomposes to its
constituent nucleons (breakup channel), may be correlated with either
a nuclear mean-field or nucleonic medium effect on vector-meson
properties (Section~\ref{sec:int}).  Specifically, exclusive
subthreshold photoproduction produces $\rho^0$ mesons which, on the
average, traverse distances comparable to the size of the nucleus
before their hadronic decay. This process is therefore a probe of
vector-meson medium modifications at normal nuclear densities, a
regime which has been the focus of mean-field driven theoretical
models. The breakup channel, on the other hand, is more likely to
produce $\rho^0$'s which are slower relative to the struck nucleon and
may travel distances shorteer than the nucleonic radius before decaying
(Section~\ref{sec:int}). Thus, large densities in the interior of the
nucleon may become accessible via the breakup channel, amounting to a
nucleonic medium effect. This is the realm of the emerging
hadronic-quark nature of matter, a domain virtually unexplored. In the
deep subthreshold region, and for large momentum transfers to the
target nucleus, the coherent and exclusive-quasifree channels are
suppressed, and the breakup-quasifree process becomes dominant.  It is
the subthreshold dynamics that motivated the 800$\le$E$_\gamma\le$1120
MeV $\rho^0$ photoproduction experiment.

The aim of the experiment being as stated above, the choice of
${^3}He$ as the target becomes almost ideal. The low photon energies
utilized to induce subthreshold photoproduction result in slow
$\rho^0$ mesons with a small Lorentz boost, and therefore a large
probability for decay within the nuclear volume.  Thus, in the case of
the exclusive channel, a large nuclear radius is not necessary, and
the larger nuclear density of a heavier target is predicted to have
only a marginally enhanced effect on the $\rho^0$ meson mass
\cite{sai}.  Furthermore, if the breakup process is dominated by the
nucleonic field, the size of the target nucleus is
irrelevant. Finally, the ${^3}He$ target is the lightest nucleus where
a nuclear medium effect may be discernible, without the complexity of
overwhelming final state interactions (FSI).

The $\rho^0$ detection is further aided by the inherent selectivity of
the TAGX spectrometer (Section~\ref{sec:spec}) to coplanar
$\pi^+\pi^-$ processes \cite{maru}. This is due to the limited
out-of-plane acceptance of the spectrometer, which preferentially
selects the $\rho^0 \rightarrow \pi^+\pi^-$ channel, at the expense of
two-step processes (e.g. FSI) and uncorrelated $\pi^+\pi^-$ production
at distinct reaction vertices, the latter accounting for the majority
of non-$\rho^0$ background events (Section~\ref{sec:pcs}). This
favorable feature of the spectrometer promotes an otherwise small
component of the total amplitude, namely the subthreshold breakup
channel, to a sizeable experimental signal.

The mass of the $\rho^0$ meson in the nuclear medium was extracted
from the dipion spectra of the 800$\le$E$_\gamma\le$1120 MeV
experiment with the aid of Monte Carlo (MC) simulations
\cite{lolos}. The reported mass shift was far larger than the
predictions of any mean-field driven model pertaining to the exclusive
process for ${^3}He$ \cite{sai}; a calculation based on QHD assuming a
deep scalar potential yielding a $\rho^0$-${^3}He$ bound state, on the
other hand, produced much better agreement\cite{zisis}. The result of
Ref.~\cite{lolos} led to a reanalysis of the lower-energy measurements
\cite{watts} including the $\rho^0 \rightarrow \pi^+\pi^-$ channel,
and allowing for $\rho^0$ production with a shifted mass. The outcome
of this reanalysis was an even larger shift \cite{hub}, possibly
indicating a mechanism other than those previously considered. A
nucleonic medium effect, as sketched earlier, may be consistent with a
large $\rho^0$-mass reduction, although a theoretical model has yet to
be fully developed \cite{geb4,guic}. Though more work is needed to
firmly establish and better understand these results \cite{sp}, the
${^3}He$($\gamma$,$\pi^+\pi^-$)X experiment for photon energies in the
range 800$\le$E$_\gamma\le$1120 MeV constitutes the first direct
measurement of the $\rho^0$ mass in the nuclear medium. In this
report, a new and more thorough analysis is presented, including the
first direct evidence of the characteristic $J = 1$ signature of the
$\rho^0$ meson decay in the subthreshold region, as well as
refinements in the simulations and fitting procedure, relative to the
analysis of Refs.~\cite{watts,lolos,hub}, leading to higher confidence
in the extraction of the in-medium $\rho^0$ mass.

The paper is organized in six sections. In Section~\ref{sec:tagx},
the experimental set-up and the calibration procedure are reviewed. In
Section~\ref{sec:anal} the data-analysis algorithm is outlined in
conjunction with the experimental aspects of
Section~\ref{sec:tagx}. Section~\ref{sec:mc} focuses on the MC
techniques and the fitting of the data, leading to the extraction of
the mass shift. In Section~\ref{sec:data} the data are compared
with the MC calculations, and the $\rho^0$ mass shifts are discussed.
Finally, the conclusions are presented in Section~\ref{sec:concl}.

\section{APPARATUS}
\label{sec:tagx}

The INS tagged photon beam and the TAGX spectrometer (Fig.~\ref{figtgx})
\cite{maru}, the new straw drift chamber \cite{gar}, and different
aspects of the data-analysis procedure \cite{maru,watts} have all been
described in detail elsewhere, where the reader is referred for a more
extensive discussion. In this section, a brief overview of the
experimental apparatus is provided, as it applies for the
${^3}He$($\gamma$,$\pi^+\pi^-$)X measurements, stressing the elements
that were either introduced for the first time in this experiment, or
that are important for the data analysis.

\subsection{Photon beam and ${^3}He$ target}
\label{sec:pb}

The photon beam is produced utilizing the 1.3 GeV Tokyo Electron
Synchrotron. A series of innovative technical improvements led in 1987
to the upgrading of the photon beam to one of medium duty cycle. In
the present experiment, the endpoint electron energy E$_s$ is 1.22 GeV
at an average 10\% duty factor, corresponding to a 5 ms extraction
time. The instantaneous energy of the extracted electrons has a
sinusoidal dependence, and it is known by measuring the extraction
time. The extracted electrons are directed via a beamline onto a thin
platinum radiator where bremsstrahlung photons are produced, while the
scattered electrons are bent away from the beam by a rectangular
analyzer magnet of 1.17 T. The magnet settings of the extraction
beamline vary sinusoidally in time, in phase with the E$_s$ energy.
An array of 32 scintillator electron-tagging counters, each with a 10
MeV/c momentum acceptance, detect the scattered electrons. The
position of the tagger registering the scattered electron determines
its energy, and consequently that of the bremsstrahlung photon as
E$_\gamma$ = E$_e$ - E$_{e'}$ with $\Delta$E$_\gamma\sim \pm$5~MeV
accuracy. A second set of 8 backing taggers participates in the
coincidence triggering signal, along with the 32 frontal ones, and
is discussed in the following section. The tagged photon intensity was
maintained at an average of $\sim$3.5 $\times$ 10$^5$ $\gamma$/s, well
within the tolerance of the data acquisition system for accidental
triggers.  The photons, distributed over a beam spot of $\sim$2~cm in
diameter, are subsequently incident on a liquid ${^3}He$ target. The
target is at a temperature of 1.986$\pm$0.001 K, corresponding to a
0.0786 g/cm$^3$ density, and is contained in a cylindrical vessel 90
mm in height and 50 mm in diameter at the center of the TAGX
spectrometer \cite{target}.

The tagger hits are related to the photon flux incident upon the
target after efficiency corrections. In particular, due to the
collimation of the photon beam downstream from the taggers, some of
the tagged photons do not reach the target. To determine the tagging
efficiency, and consequently the photon flux, a lead-glass
$\breve{C}$erenkov counter is placed in the photon beam, with reduced
flux, downstream from the target and the tagger scalers are
periodically recorded both with and without the platinum radiator in
place.  The efficiency per tagger counter is determined by the
relation
\begin{equation}
\eta_{\imath\in[1,32]}={ {\left[\breve{C}\cdot T_\imath\right]_{in}
- \left[\breve{C}\cdot T_\imath\right]_{out}
- \left[\breve{C}_{acc}\cdot T_\imath\right]_{in}
}\over{T_{\imath_{in}}
- T_{\imath_{out}}} }
\label{eta}\end{equation}
where $T_\imath$ is the scaler count for each of the frontal taggers.
The term $\left[\breve{C}\cdot T_\imath\right]_{out}$ corresponds to
the coincidences between a tagger-counter and a lead-glass
$\breve{C}$erenkov hit from a spurious photon not originating from the
platinum radiator, and $\left[\breve{C}_{acc}\cdot
T_\imath\right]_{in}$ is the number of accidental coincidences with
the radiator in place, but with the $\breve{C}$erenkov hit registering
with a delay of the order of 100~ns with respect to the tagger counter
signal. These two terms turn out to be negligible relative to the term
$\left[\breve{C}\cdot T_\imath\right]_{in}$ in Eq.~(\ref{eta}), which
gives the efficiency-corrected number of tagged photons per tagger
reaching the target as
\begin{equation}
N_{\imath\in[1,32]} = \eta_\imath T_{\imath_{in}} \left(1-{
{T_{\imath_{out}}\over{T_{\imath_{in}}}}}\right) \ .
\label{norm}\end{equation}
The efficiencies $\eta_\imath$ and radiator out/in ratios ${\cal
R_\imath}$ = $T_{\imath_{out}}/T_{\imath_{in}}$ for the 32 frontal
taggers are recorded in a number of dedicated runs, in regular
intervals throughout the experiment, and they are used, along with
Eq.~(\ref{norm}), in the empty-target background subtraction procedure
applied to the measured spectra (see Section~III).

\subsection{Spectrometer}
\label{sec:spec}

The TAGX spectrometer has an acceptance of $\pi$ sr for charged
particles (neutral-particle detection was not utilized in the present
experiment), and has been in use at INS since 1987. It consists of
several layers of detector elements (Fig.~\ref{figtgx}) positioned
radially outwards from the target vessel, which is located at the
center of the 0.5~T field of a dipole analyzer magnet.

Directly surrounding the target container is the inner hodoscope (IH),
made up of two sets of six scintillator counters, one on each side of
the beam. The IH is used in the trigger, as well as in measuring the
time of flight (TOF) of the outgoing particles \cite{maru,gar}. As it
is placed inside a strong magnetic field, the light signal is carried
by optical fibers to the photomultiplier tubes, which are located at
the fringes of the magnetic field two meters away.

Next is the straw drift chamber (SDC), operating since 1994 and
installed expressly for the measurement of the $\rho^0$ mass in
${^3}He$, with the objective of improving the momentum resolution for
the detection of the $\rho^0 \rightarrow \pi^+\pi^-$ decay channel
\cite{gar}. The SDC consists of two semi-circular cylindrical
sections, each containing four layers of vertical cells. The ``straw''
cells have tube-shaped cathodes which induce a radial electric field,
and consequently have a regular field definition and high position
resolution ($\sim$150~$\mu$m). The SDC was designed to preserve the
$\pi$-sr large acceptance prior to its installation, to not impose
extensive modifications of the spectrometer, and to not induce
significant energy losses to traversing particles by keeping its
thickness to minimum.  The installation of the SDC required,
nonetheless, the replacement of the IH from an earlier set of
scintillators with the one described above.

Surrounding the SDC are two semicircular cylindrical drift chambers
(CDC) subtending angles from 15$^\circ$ to 165$^\circ$ on both sides
of the beamline in the horizontal plane, and $\pm$18.3$^\circ$ in
vertical out-of-plane angles. The CDC is composed of twelve concentric
layers of drift cells, yielding a $\sim$250-300~$\mu$m horizontal and
$1.5$~cm vertical resolution. Together with the SDC, they are used to
determine the planar momentum and emission angle of the charged
particles traversing them, and the vertex position of trajectory
crossings.

The outer set of 33 scintillator elements comes next, serving as the
outer hodoscope (OH), with PMT's attached at both ends to help
determine the track angle relative to the median plane. The two sets
of hodoscopes, IH and OH, measure the TOF corresponding to the tracked
trajectories.

Other components of the TAGX spectrometer include 4 sets of
155~mm~$\times$~50~mm~$\times$~5~mm $e^+e^-$ background veto counters
positioned along the OH arms in the median plane. The veto counters
eliminate charged-particle tracks registering within $\Delta z = \pm$
2.5 mm, mostly affecting forward-focused $e^+e^-$ pairs produced
copiously downstream from the target, but having a small effect on
$\pi^+\pi^-$ events.

\subsection{Data acquisition and calibration}
\label{sec:daq}

The channel of interest being $\pi^+\pi^-$ production from the decay
of the $\rho^0$ meson, two-charged particle coincidences on opposite
sides of the beam axis were required of the trigger. Two levels of
triggering are implemented in order to optimize the data acquisition
electronics. The pretrigger
\begin{equation}
{\cal PT}=IH_L\cdot IH_R \cdot \sum_{\imath=1}^8 TAG_{\imath_{back}} \cdot
{\overline {EM}}_{front}
\label{ptrig}
\end{equation}
is generated within 100~ns from the occurence of an event. A
coincidence of a left and right IH hit with a backing tagger hit is
required, and not rejected by the forward $e^+e^-$ veto counters. The
main trigger
\begin{equation}
{\cal MT}={\cal PT} \cdot OH_L\cdot OH_R \cdot \sum_{\imath=1}^{32}
TAG_{\imath_{front}} \cdot
{\overline{Inhibit}}
\label{mtrig}
\end{equation}
requires the coincidence of the pretrigger with a left and right OH
hit and a forward tagger hit, not rejected by the computer inhibit
signal. A window of 400~ns is available between the ${\cal PT}$ and
the ${\cal MT}$, after which the CAMAC is cleared for the next ${\cal
PT}$. Typical counting rates are 2~kHz and 30~Hz for the ${\cal PT}$
and ${\cal MT}$, respectively.

The calibration of the scintillation counters and the CDC and SDC have
been the subject of extensive effort \cite{maru,gar,iur}. More
recently, a series of modifications implemented in the track fitting
algorithms has resulted in significant improvements, mainly in the
planar-momentum resolution \cite{aki}.  It is the tracking that is
discussed next.

\begin{figure}
\epsfig{file=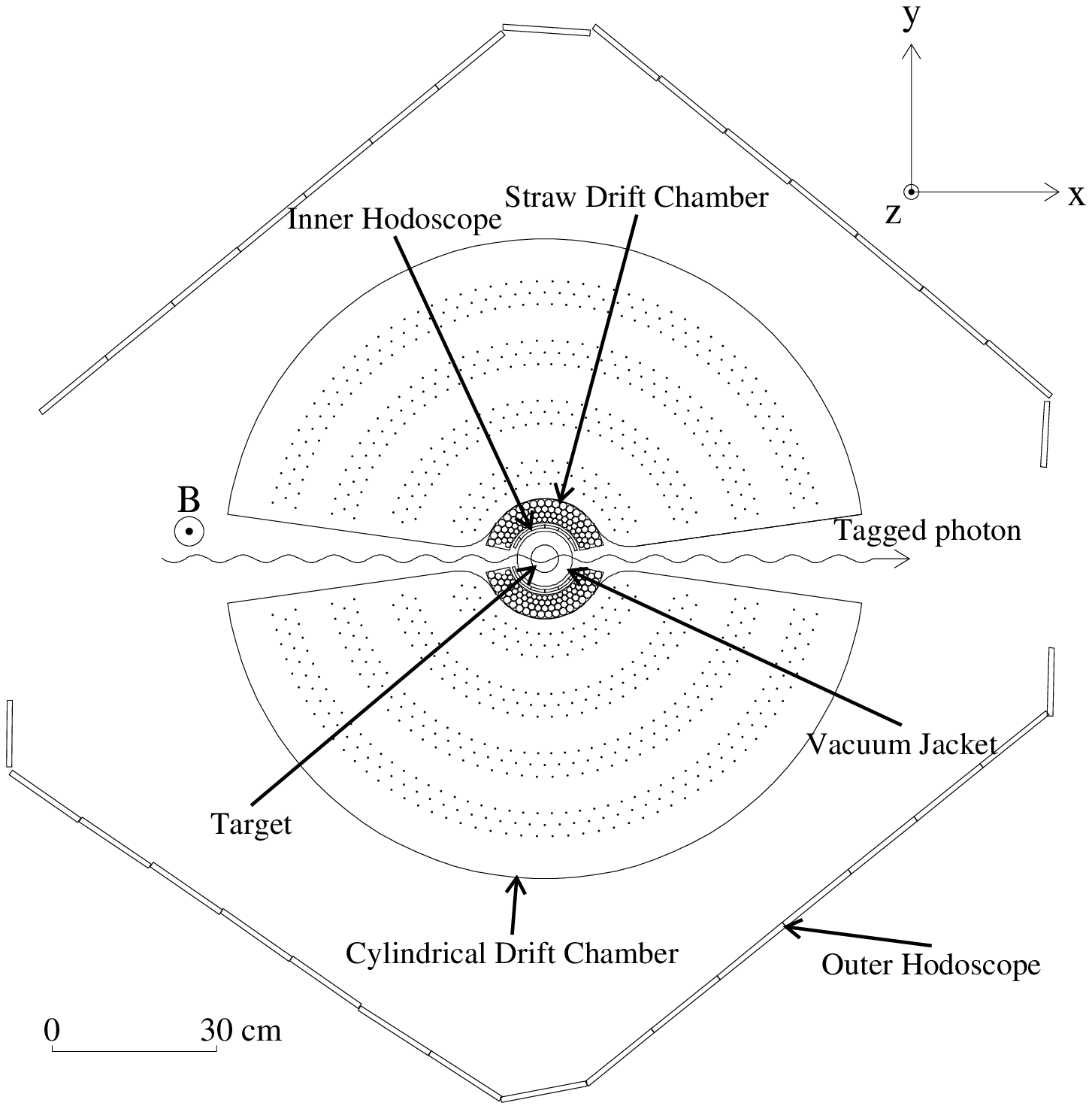}
\caption{Components of the TAGX spectrometer: Radially outwards from
the target vessel at the center are seen the IH, SDC, CDC, and OH. The
TAGX coordinate frame has the ${\hat x}$-axis pointing in the
direction of the photon beam, and the ${\hat z}$-axis pointing outward
from the page in the direction of the 0.5~T magnetic field of the
dipole analyzer magnet.  Veto counters (not shown in the figure) are
positioned along the OH arms in the xy-plane, for $e^+e^-$-pair
rejection, mainly at forward angles.}
\label{figtgx}
\end{figure}

\begin{figure}
\epsfig{file=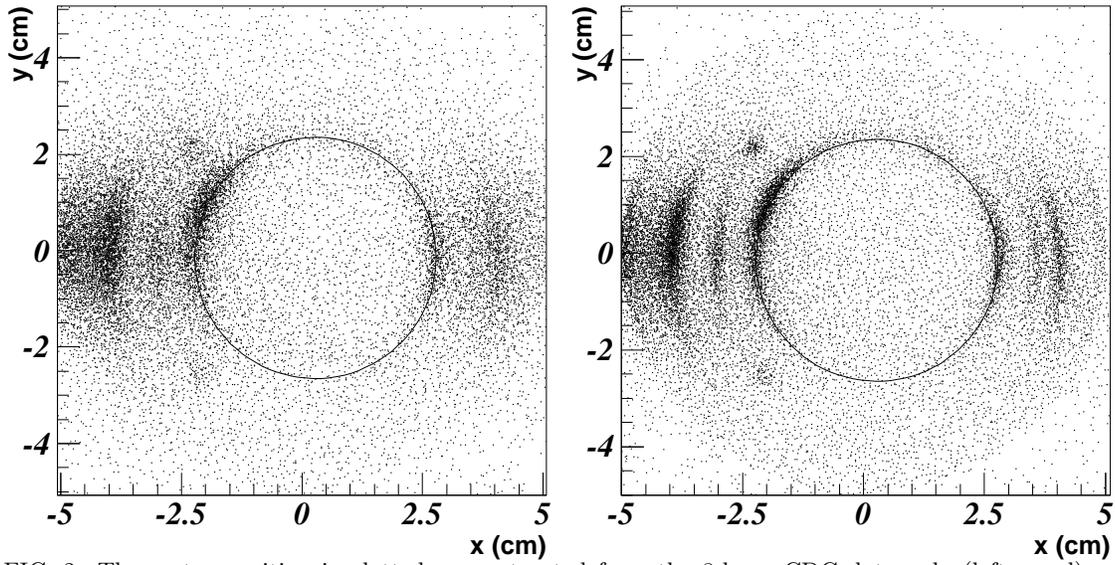}
\caption{The vertex position is plotted, reconstructed from the
8-layer CDC data only (left panel), and from the 9-layer CDC and SDC
data (right panel), for a set of empty-target runs
\protect{\cite{aki}}. The evident improvement in the resolution is
discussed in the text. The components of the target vessel, visible in
the figure, have been discussed elsewhere \protect{\cite{gar,target}}.
The inner ring corresponds to the Mylar wall containing the liquid
${^3}He$ target, a cylinder 50~mm in diameter. A circle indicating the
position of the target container has been drawn, centered at
(x,y)=(2.8~mm,-1.5~mm) in the TAGX coordinate frame.}
\label{figsdc}
\end{figure}

The CDC consists of four groups of three-wire layers
(Fig.~\ref{figtgx}).  The last layer of wires for each group was
intended for charge division readout and had not been employed in the
past \cite{maru}.  Instead, hits from the eight remaining CDC layers
were used for the reconstruction of the planar momentum $p_{xy}$,
emission angle $\theta_{sc}$, and vertex position (see also
Section~\ref{sec:obs}). This earlier eight-layer tracking procedure
did not incorporate the SDC information either, thus resulting
altogether in a less-than-optimal momentum resolution.  Since longer
effective lengths of reconstructed tracks result in higher-quality
fits, however, TDCs from the last layer of wires of the fourth group
has been implemented for the first time in the present analysis.
Furthermore, the SDC data have also been used for the first time, a
combination which yields an overall improvement in the planar momentum
resolution estimated to be $\sigma_{p_{xy}}$/p$_{xy}$ = 0.0892p$_{xy}$
+ 0.0057, compared with $\sigma_{p_{xy}}$/p$_{xy}$ = 0.1150p$_{xy}$ +
0.0078 from the 8-layer CDC analysis \cite{aki}. For a particle of
p$_{xy}$=300~MeV/c, this amounts to a 40\% improvement in the planar
momentum resolution.  The corresponding improvement in the vertex
position is reflected in Fig.~\ref{figsdc}, and has allowed for more
stringent tests in the selection of two-track events which originate
from the target area. A minor improvement has also been noted in the
emission angle resolution, which stays relatively constant at
$\sigma_\varphi \sim$~0.3$^\circ$ throughout the range of typical
planar momenta 100 MeV/c $\le$ p$_{xy} \le$~500~MeV/c \cite{aki}.

The steps involved in the tracking of trajectories through the SDC and
CDC may be summarized in the following: The CDC TDCs operate in
``common-stop'' mode, with the start determined from each CDC sense
wire and the stop from the IH \cite{maru}. The CDC drift times are
first corrected by the corresponding TDC timing offsets. The
drift-length to drift-time relation is determined next, per layer of
CDC wires, as a fifth-order polynomial.  This is an iterative process,
where an initial set of parameters is used to reconstruct a sample of
well-defined high-momentum tracks.  The reconstructed trajectories
yield a new set of parameters, and the procedure is repeated until the
convergence condition is reached, namely that the residual root mean
square (RRMS) improvement over the final two cycles is no better than
0.5\%.

Once a CDC track has been reconstructed, a similar procedure is
followed for the SDC, where first the TDC timing offset is determined,
and subsequently a SDC length-to-time relationship is extracted. This
accomplished, ``best'' SDC tracks are identified, which qualify as
candidate extensions of a selected CDC track. Typically, 2-4 SDC
tracks are selected as possible extensions of a reconstructed CDC
track when all four SDC layers have registered a hit, to a maximum of
8 candidate tracks if one SDC layer is missed. The SDC tracks are
approximated by straight-line segments, since the error in the
position of even the slowest particles which may be expected to result
in valid two-track events is within the 150~$ \mu$m tolerance of the
SDC.  Finally, the SDC candidates are matched with the CDC
reconstructed track, by requiring the minimal CDC + SDC RRMS of the
combined track.

The obtained TOF resolution $\sigma_t$ is better than 380~ps
\cite{maru}.  The TOF is used for particle identification, as well as
for the determination of the particles' OH position (along the ${\hat
z}$-axis in the TAGX frame as shown in Fig.~\ref{figtgx}).

\section{DATA ANALYSIS}
\label{sec:anal}

The data presented in this report were collected in two periods, with
${^3}He$ and empty-target measurements in each. The superior quality
of the photon beam, and a longer running period, resulted in better
statistics and a higher ratio of $\pi^+\pi^-$ to accidental triggers
for the second phase. This is reflected in the tagger efficiencies and
radiator out/in ratios, defined in Eq.~(\ref{norm}), which during the
later part of the experiment were generally improved. A total of
16,366 $\pi^+\pi^-$ events have been identified from the analysis of
two-track events, comprising 73\% of the total number of reconstructed
events, the remaining being of three (23\%) or more tracks
($<$4\%). With the extraction of the $\pi^+\pi^-$ yield Y, the total
cross section $\sigma_T$ is determined from the relation
\begin{equation}
\sigma_T = {Y\over{N_TN_\gamma\eta_{\pi^+\pi^-}\eta_{daq}}}
\label{yield}
\end{equation}
where $N_T$ (nuclei/cm$^2$) is the ${^3}He$ target density seen by the
photon beam, and $N_\gamma$ is the incident photon flux.  The
efficiencies $\eta_{daq}$ and $\eta_{\pi^+\pi^-}$ account for the
data-acquisition livetime and $\pi^+\pi^-$ detection efficiencies.
The latter, which is in the range of 2.7-6.8\% for the $\rho^0$
channels, is determined by dedicated MC routines and is
reaction-channel specific (see Ref.~\cite{maru}).

\subsection{Empty target background}
\label{sec:ets}

In Section~\ref{sec:pb}, the extraction of the tagger efficiencies,
and the normalization of the tagger scalers to reflect the number of
photons incident on the target, were discussed. These are utilized in
determining the appropriate factor by which empty-target spectra are
scaled prior to their subtraction from the corresponding
${^3}He$-target spectra, for the removal of target background
counts. The procedure is briefly summarized in the following steps.

\begin{figure}
\epsfig{file=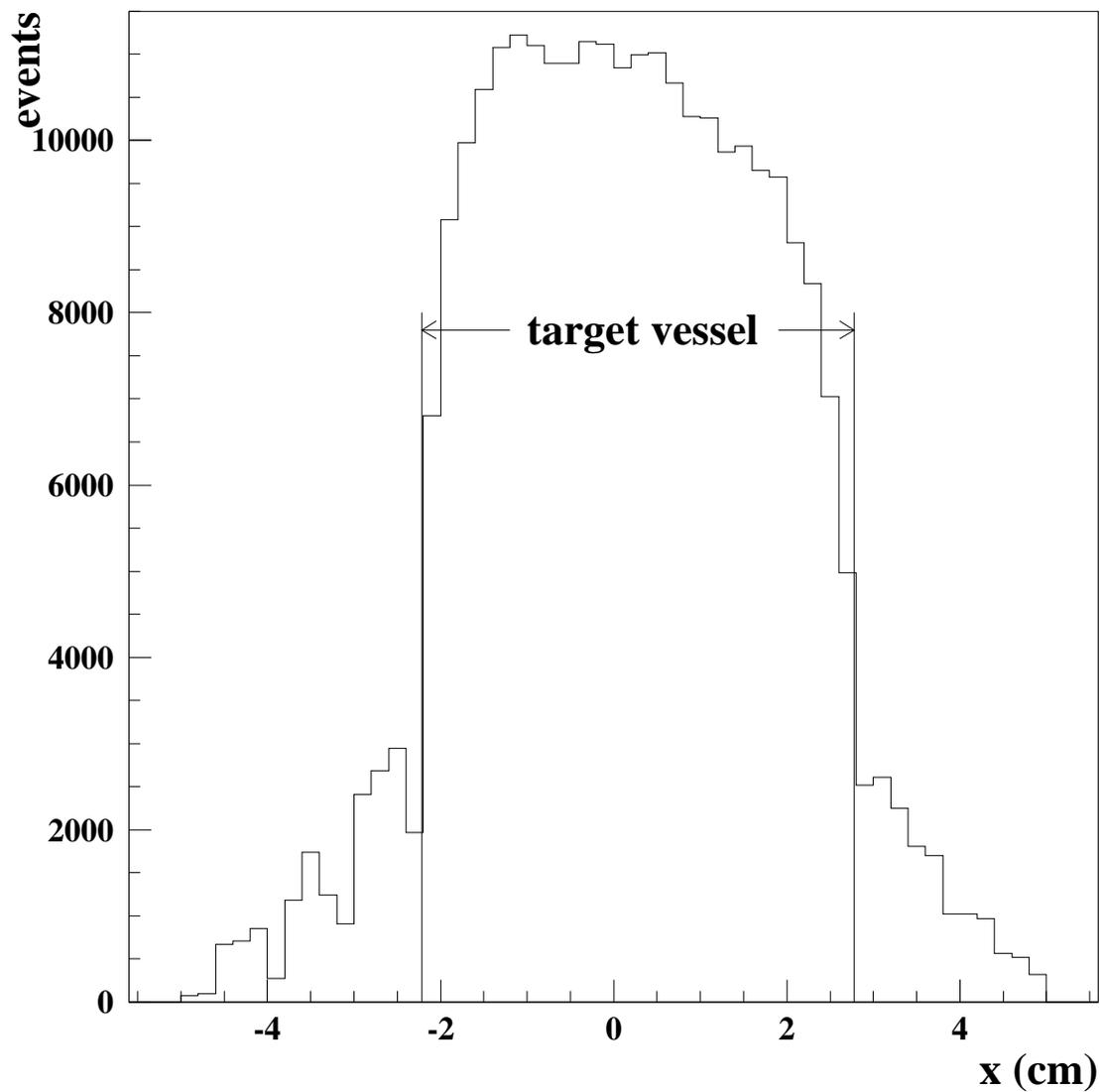}
\caption{The x-coordinate of the vertex position after background
subtraction is shown, along with the target-vessel walls indicated as
lines. Only $\pi^+\pi^-$ events from this target region are considered
in the analysis.}
\label{figtv}
\end{figure}

At regular intervals throughout each of the ${^3}He$-target and
empty-target running periods, the lead-glass $\breve{C}$erenkov
counter is employed in dedicated reference runs to determine the
quantities $\eta_\imath$ and ${\cal R}_\imath$, as described in
Section~\ref{sec:pb}. The total number of photons incident on the
target per experiment is extracted as the sum of the raw scaler counts
T$_\imath$, recorded for each run, corrected by the efficiency and
out/in ratios for that run, according to Eq.~(\ref{norm}), and
weighted by the normalized energy distribution of the scattered
electrons. In particular, $\eta_\imath$ and ${\cal R}_\imath$ for each
run are calculated from the corresponding quantities of the reference
runs, on the assumption that they vary linearly with the raw scaler
counts accumulated between runs. The ratios of photon fluxes between
each ${^3}He$-target and its corresponding empty-target experiment
yield the scaling factors by which the latter are normalized prior to
subtraction from the former.  The x-coordinate spectrum of the
two-pion crossing vertex after background subtraction, indicative of
the accuracy of this procedure, is shown in Fig.~\ref{figtv}.

\subsection{Experimental observables}
\label{sec:obs}

\begin{figure}
\epsfig{file=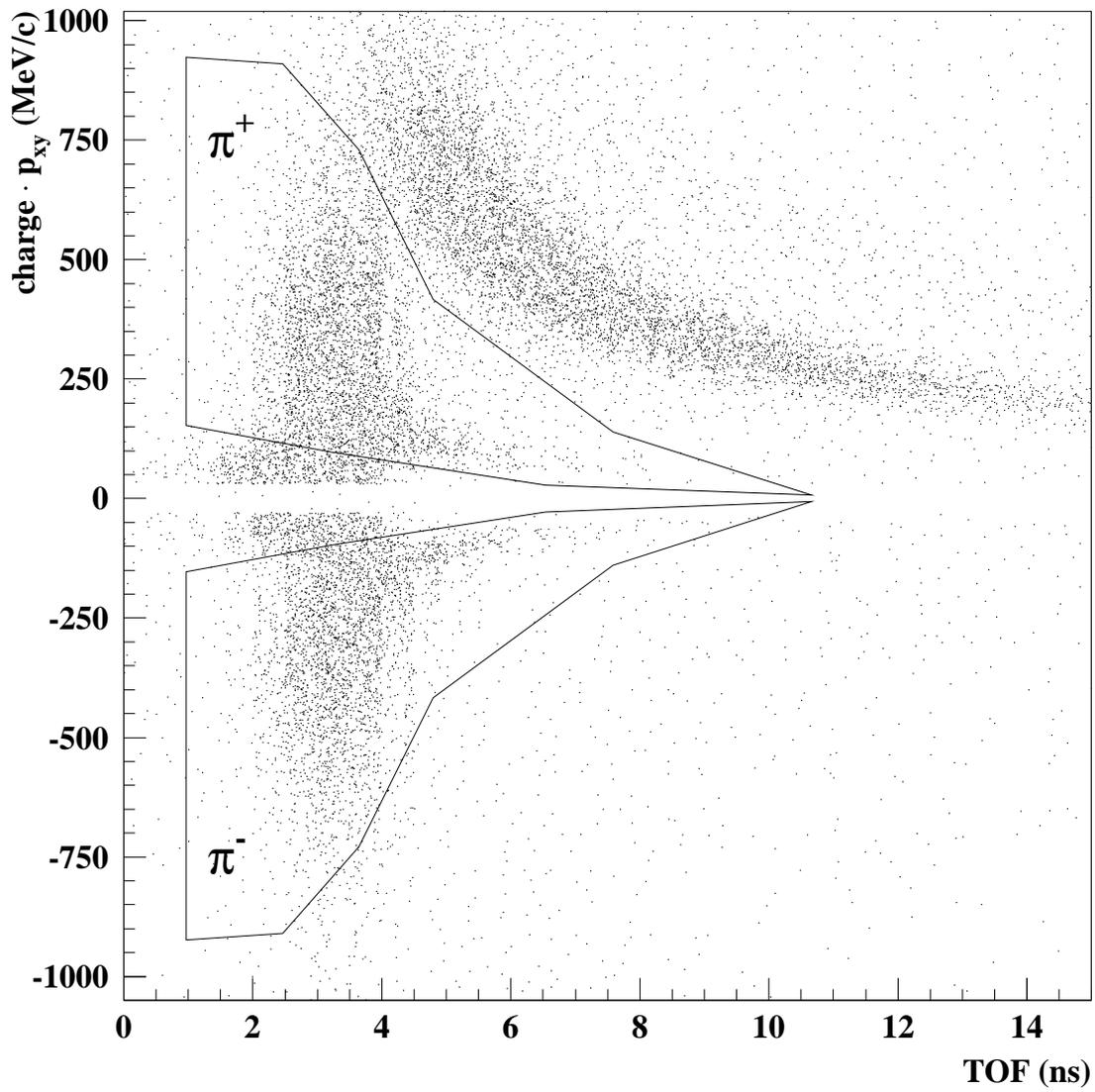}
\caption{The TOF-vs-planar momentum spectra, displaying proton
(upper-most), $\pi^\pm$ (in the box cuts), and $e^\pm$ (adjacent to
the p$_{xy}$ = 0 axis) bands, are used for particle identification.}
\label{figtvm}
\end{figure}

\begin{figure}
\epsfig{file=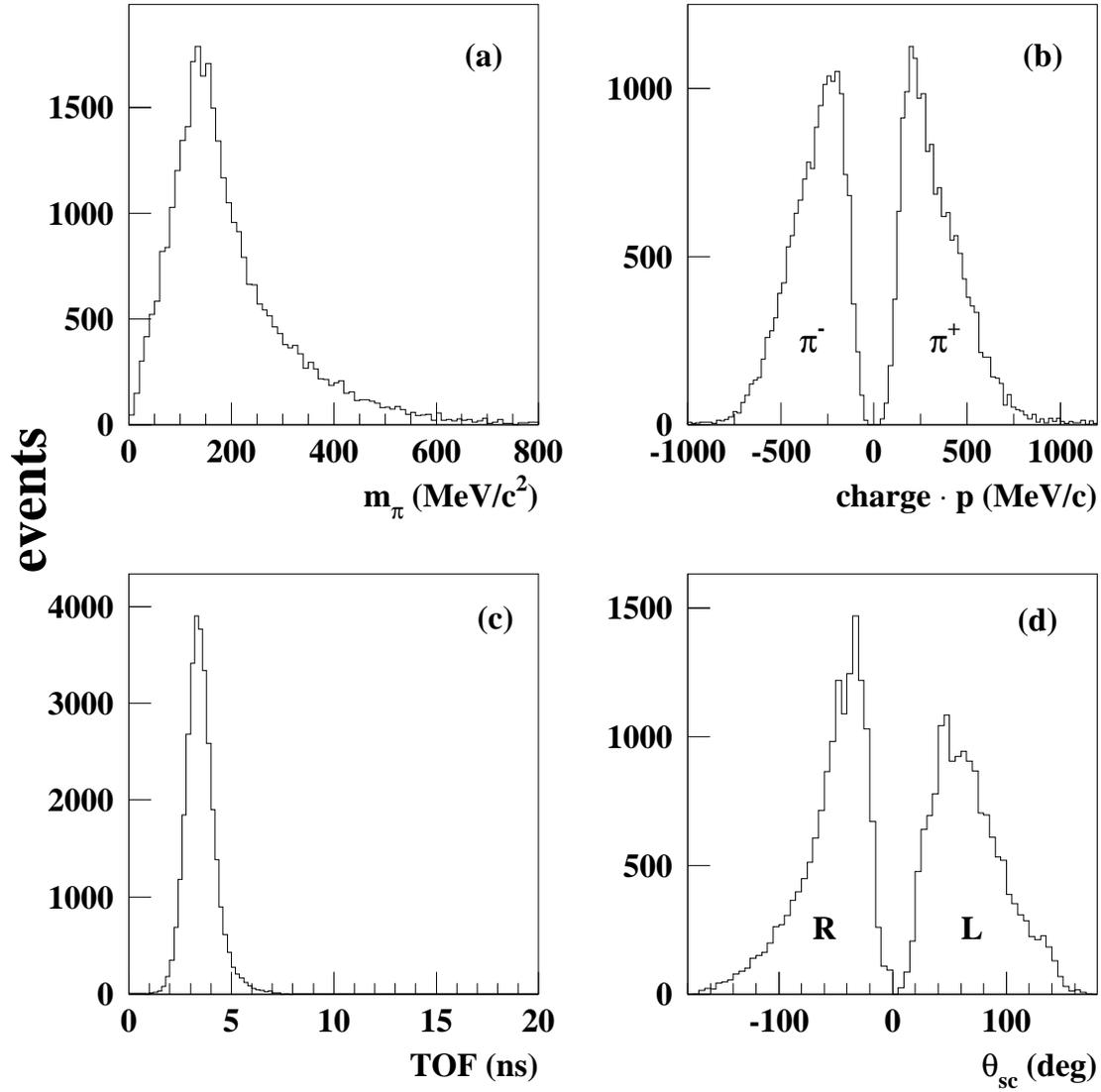}
\caption{Some of the kinematical observables for $\pi^+\pi^-$ events,
described in the text, are shown.  The sign of the momentum coresponds
to the pionic charge. The sign of the scattering angle $\theta_{sc}$
depends on whether the track registered to the left ($>$0) or right
($<$0) of the beam (see Fig.~\protect{\ref{figtgx}}).}
\label{figkin}
\end{figure}

The calibration procedure, discussed in Section~\ref{sec:daq}, allows
the extraction of the planar momentum $p_{xy}$, the polar emission
angle in the median plane $\varphi$, the planar trajectory length
$l_{xy}$ from the SDC + CDC particle tracking, and the time of flight
and z-coordinate (OH position) from the IH and OH scintillators:
\begin{eqnarray}
&&t = {1\over 2}\left(t^{up}_{OH}+t^{down}_{OH}\right) - t_{IH}\nonumber \\
&&z = {1\over 2}\left(t^{down}_{OH}-t^{up}_{OH}\right)v_{eff}
\label{ihoh}
\end{eqnarray}
The up-down indices correspond to the timing measurements at the two
ends of the OH, and $v_{eff}$ is the effective light transmission
velocity in the scintillator material. These yield the primary
observables (Figs.~\ref{figtvm},\ref{figkin})
\begin{eqnarray}
&&\theta_{dip} = tan^{-1}\left({z\over l_{xy}} \right)\nonumber \\
&&p = p_{xy}/cos\,\theta_{dip}\nonumber \\
&&l = l_{xy}/cos\,\theta_{dip}\nonumber \\
&&\beta = l/ct\nonumber \\
&&m = p/\beta\gamma c\nonumber \\
&&\theta_{sc} = cos^{-1}\left( cos\varphi\,{p_{xy}\over p} \right)
\label{pkin}
\end{eqnarray}
where $\theta_{dip}$ is the out-of-plane dip angle, $p$ and $l$ the
total momentum amplitude and three-dimensional trajectory length, and
$\theta_{sc}$ the scattering angle with respect to the incident beam
(Fig.~\ref{figkin}).  A left-right asymmetry noted in the scattering
angle $\theta_{sc}$ spectrum (Fig.~\ref{figkin}d) has been reproduced
in the MC simulations.

A coincidence of two charged particles, one on either side of the
photon beam, signifies the occurence of an event
(Section~\ref{sec:daq}). A series of tests and cuts in the data set
subsequently eliminate all but the $\pi^+\pi^-$ pairs. In particular,
first the time-of-flight versus planar momentum spectra are used for
particle identification (PID, Fig.~\ref{figtvm}). The great majority
of events including a proton or e$^\pm$ are thus discarded. Cuts on
the tagger TDC spectra reject events induced by spurious photons, not
corresponding in timing with the beam pulse. Last, pairs with
low-confidence tracks (large RRMS), or whose vertex falls outside the
target area (Fig.~\ref{figsdc}) are eliminated, thus completing a
first-level selection based on directly measured observables.

For the two-track events that have cleared the tests above, and have
been identified as $\pi^+\pi^-$, additional kinematical observables
are calculated.  At this stage, the few surviving two-track events
involving a proton or e$^\pm$ that had been previously misidentified
as $\pi^+\pi^-$ by the PID cuts are eliminated as well. The calculated
observables include the dipion invariant mass m$_{\pi^+\pi^-}$, the
laboratory production angle of the dipion system $\Theta_{IM}$, the
missing mass m$_{miss}$ and momentum p$_{miss}$, the $\pi^+$-$\pi^-$
laboratory opening angle $\vartheta_{\pi^+\pi^-}$, and the $\pi^+$
production angle in the dipion center of mass $\theta^*_{\pi^+}$,
employed as variables in the MC fitting procedure
(Section~\ref{sec:mc}).

Among these observables, the production angle for either one of the
two pions in the dipion center-of-mass frame, for example
$\theta^*_{\pi^+}$, is singular as a direct experimental observable
which, without the aid of simulations or assumptions, points to the
presence of the $\rho^0$ production channel well below the nominal
threshold energy. This is discussed next.

\subsection{The $J=1$ signature of the $\rho^0$}
\label{sec:le1}

Among the dominant production channels participating in $\pi^+\pi^-$
photoproduction in the E$_\gamma \sim$ 1 GeV region
(Section~\ref{sec:pcs}), the $\rho^0 \rightarrow \pi^+\pi^-$ channel
alone results in the two pions being produced at a single reaction
vertex with the characteristic $J = 1$ angular correlation from the
decay of the $\rho^0$. In the dipion center-of-mass frame this
translates into a pure cos$^2\theta^*_{\pi^+}$ distribution, where
$\theta^*_{\pi^+}$ is the $\pi^+$ production angle with respect to the
dipion momentum, the direction defined by the latter in the laboratory
frame. On the assumption of a slowly-varying $\pi^+\pi^-$ background
interfering with the $\rho^0$ amplitude, the angular distributions are
expected to be symmetric around $\theta^*_{\pi^+}$ = 90$^\circ$ for
dipion cm energies near the mass of the $\rho^0$ meson, where the
$\rho^0 \rightarrow \pi^+\pi^-$ amplitude peaks. Away from the
$\rho^0$ mass, the resonant amplitude vanishes, and the background
processes dictate the shape of the spectra \cite{rhob}. Thus, above
and below the $\rho^0$-meson mass, the angular distribution is
expected to regain a quasi-isotropic shape due either to the
uncorrelated pions produced at two or more reaction vertices, this
being the case for the majority of the participating background
processes, or from s-wave correlated pions, possibly produced from the
decay of the $\sigma$ meson (Section~\ref{sec:pcs}).

This technique, of $\rho^{0,\pm}$ identification via the study of the
pion-scattering angle in the dipion cm frame, has been extensively
used in many previous analyses (e.g. Ref.~\cite{rhoa}). The
$\theta^*_{\pi^+}$ distribution spectrum, based on the above, is
expected to be well-described as A + B\,cos$^2\theta^*_{\pi^+}$, in
the vicinity of the dipion invariant mass matching that of the
$\rho^0$ meson, where the $\rho^0 \rightarrow \pi^+\pi^-$ amplitude
peaks.

The $\pi^+\pi^-$ events in the range of 400-800 MeV/c$^2$ in dipion
invariant mass have been divided in four 100 MeV/c$^2$ bins, which is
the finest binning allowed by the data statistics. An additional cut,
determined from the MC simulations of the TAGX $\rho^0$ detection
efficiency and kinematical considerations, eliminates those
$\pi^+\pi^-$ events with too small an opening angle to have been the
outcome of back-to-back production at a single reaction vertex (see
Section~\ref{sec:fp}). The latter cut results in a further 9-10\%
reduction in the total number of $\pi^+\pi^-$ events in the 400-800
MeV/c$^2$ dipion invariant-mass region, affecting only events from
background processes, effectively boosting the $\rho^0 \rightarrow
\pi^+\pi^-$ amplitude relative to the background (Fig.~\ref{figcos}).

The 500-600, and 600-700 MeV/c$^2$ regions (Figs.~\ref{figcos}b,c)
clearly demonstrate the $J=1$ fingerprint of the $\rho^0$ meson
decay. The deviation from cos$^2\theta^*_{\pi^+}$ toward $\pm$1 is
reproduced in MC simulations of the $\rho^0 \rightarrow \pi^+\pi^-$
process, and it is shown to be the effect of the TAGX detection
efficiency, stemming from the two-track detection requirement (see
Section~\ref{sec:fp} and Fig.~\ref{figacc}). The quasi-resonant
$\rho^0 \rightarrow \pi^+\pi^-$ amplitude over the 500-700 MeV/c$^2$
dipion invariant-mass range points to a substantially reduced $\rho^0$
mass {\it beyond} the trivial apparent lowering, which is the artifact
of probing the lower tail of the $\rho^0$ mass distribution with
low-energy photons (Section~\ref{sec:int}).

\section{SIMULATIONS}
\label{sec:mc}

The MC simulations constitute an integral part of the data analysis by
determining the process-dependent detection efficiencies of the
spectrometer, guiding the assignment of the weight to each of the
contributing production mechanisms, and, ultimately, leading to the
extraction of the medium-modified $\rho^0$ mass.

The steps involved in the simulations and fitting algorithm can be
outlined as follows:
\begin{itemize}
\item Eleven individual $\pi^+\pi^-$ production channels are coded 
into MC generators (Section~\ref{sec:pcs}). These eleven processes are
considered to account for the full $\pi^+\pi^-$ photoproduction yield
in the $\gamma + {^3}He$ reaction. Twelve distributions of six
kinematical observables, with cuts aiming to separate the bound
${^3}He$ from the dissociated $ppn$ final state, are simulated for
each production channel and each of four $\Delta$E$_\gamma$ energy
bins (Section~\ref{sec:fp}). The analysis of the MC events is
identical to that of the experimental data, and yields the
process-dependent spectrometer acceptance (Section~\ref{sec:fp}).

\item The simulated spectra for nine of the above elementary processess
(Section~\ref{sec:pcs}), including background and $\rho^0$ production
channels with the nominal m$_{\rho^0}$=770 MeV/c$^2$ mass, are
combined.  The twelve spectra of each process are adjusted with a
common strength parameter within each of the four $\Delta$E$_\gamma$
bins before being added together. Subsequently, all twelve simulated
spectra are fitted simultaneously to the corresponding experimental
ones, yielding the nine strength parameters independently for each
$\Delta$E$_\gamma$ bin. From the latter, and the spectrometer
acceptances, total cross sections are extracted for each of the nine
production processes, and compared with independently established
ones. Adjustments to the starting values and fitting constraints are
made in iterative steps until satisfactory agreement is reached.

\item The procedure is repeated for all eleven production channels, 
including the addition of two more processes (Section~\ref{sec:pcs})
with a modified $\rho^0$ mass m$^*_{\rho^0}$ in the range 500-725
MeV/c$^2$, but common for both the ${^3}He$ and breakup $ppn$ final
states. The $\rho^0$ mass corresponding to the best fit for each
$\Delta$E$_\gamma$ bin is quoted in this report as the medium-modified
m$^*_{\rho^0}$ mass (Section~\ref{sec:mmm}).

\item Exploratory fits are attempted, decoupling the ${^3}He$ and breakup 
$ppn$ final states with respect to m$^*_{\rho^0}$, as well as
modifying the width $\Gamma^*_{\rho^0}$ (Section~\ref{sec:int}).
\end{itemize}

The principal aspects of this algorithm are elaborated below.

\subsection{Production channels}
\label{sec:pcs}

Several mechanisms are known to contribute to $\pi^+\pi^-$
photoproduction. Recent experiments for photon energies between 450 -
800 MeV at MAMI, using the large-acceptance spectrometer DAPHNE and
high-intensity tagged photon beams, and in the range 1 - 2.03 GeV with
the SAPHIR detector at ELSA \cite{saphir1}, have provided accurate
measurements of the reaction $\gamma p \rightarrow \pi^+\pi^- p$
\cite{saphir2,mami1,mami2}. These have motivated several theoretical
models, which concur in their interpretation of the data as
$\pi^+\pi^-$ photoproduction predominantly through the
$\pi\Delta(1232) \rightarrow \pi^+\pi^- N$, and the $N^*(1520)
\rightarrow \pi\Delta \rightarrow \pi^+\pi^- N$ channels 
\cite{ppp1,ppp2,ppp3}. In the nuclear medium, the propagators of baryonic 
resonances require renormalization, and, in addition, many-body
effects caused by pion rescattering (FSI) are known to interfere with
the lowest-order reaction mechanism of two-pion photoproduction on the
nucleon \cite{ssk}. These medium modifications affect both the
strength and the peak position of the cross-section spectra for the
various interfering channels, relative to the corresponding processes
on a free nucleon.  Nonetheless, the $\Delta$(1232) and N$^*$(1520)
resonances remain the leading channels in photon-induced reactions in
the nuclear medium, as has recently been verified by total
photoabsorption cross-section measurements on nuclei
\cite{bianchi}. In the latter, substantial contributions were also
attributed to the nucleonic excitations P$_{11}$(1440) and
S$_{11}$(1535), primarily, which largely overlap with the N$^*$(1520)
resonance in medium-modified mass and width.

The double-$\Delta$ is another channel that has been verified in
photoabsorption measurements on the deuteron \cite{dd1,dd2}, a process
that has also been modelled theoretically \cite{on}. The photon is
absorbed on two nuclei, exciting $\Delta$(1232) resonances, in the
reaction $\gamma NN \rightarrow \Delta\Delta$, which subsequently
decay to produce $\pi^+\pi^-$ pairs.

In addition, three-pion $\pi^+\pi^-\pi^0$ production, associated with
${^3}He$ disintegration, is kinematically feasible in the energy range
probed by the present experiment. However, the limited out-of-plane
detector acceptance coupled with appropriate missing-mass cuts
minimize the contributions of this mechanism. The experimental
measurements available are sparse for this process in the energy
regime of interest \cite{saphir2,saphir3,ppp}.

Other possible contributions to the background $\pi^+\pi^-$ count,
which, however, were not found to improve the quality of the fit and
are presently not included in the simulations, may come from
non-resonant three-, four-, and five-body phase space, corresponding
to the ${^3}He$ remaining intact, or breaking-up into $dp$ and $ppn$
respectively.  These multi-body phase-space processes are governed
solely by energy and momentum conservation, each with a constant
transition matrix element \cite{sod}, and, loosely speaking,
accomodate in an average sense all the remaining possible production
channels which are too weak to be individually identified.

The contributions of the mechanisms discussed so far have been
previously considered in MC simulations, in connection with TAGX
$\pi^+\pi^-$ photoproduction data \cite{watts,lolos,hub}. In
Ref.~\cite{lolos} in particular, where $\pi^+\pi^-$ photoproduction
via the $\rho^0$ channel was first considered, the background
processes
$$
\gamma + ^3He \rightarrow \left\{ \begin{array}{l} {\left.
\begin{array}{l}
{i)\ \ \ \Delta\pi(NN)_{sp}}\\
{ii)\ \ N^*(1520)(NN)_{sp} \rightarrow \Delta(1232)\pi(NN)_{sp}}\\
{iii)\ N^*(1520)\pi(NN)_{sp}}\\
{iv)\ \ \Delta\Delta N_{sp}} \end{array} \right\} \rightarrow ppn\pi^+\pi^-}\\
\ {v)\ \ \ ppn\pi^+\pi^-\pi^0}
\end{array} \right. \,
$$
were included in simulations of non-$\rho^0$ $\pi^+\pi^-$
contributions, as well as final-state interactions (FSI) following the
$\rho^0$ decay (see process $ix)$ below). The index $sp$ signifies
spectator nucleons. The empirical values from Ref.~\cite{bianchi} were
used for the $\Delta$(1232) mass and width and for the N$^*$(1520)
mass, but the fit improved with the N$^*$(1520) width doubled relative
to Ref.~\cite{bianchi}. This {\it ad hoc} increase effectively
incorporates the near-by resonances P$_{11}$(1440) and S$_{11}$(1535),
which largely overlap with the N$^*$(1520), but cannot be resolved
within the sensitivity of the data. Alternate fits were performed with
the N$^*$(1520) replaced by the Roper N$^*$(1440) and including
five-body phase space. The two methods yield comparable masses for the
$\rho^0$, but the former is preferred as it results in a better fit.

Additional improvements in the fitting procedure, relative to the
analysis of Refs.~\cite{watts,lolos,hub}, include the modification of
the MC generators to account for the angular momentum of all $\rho^0
\rightarrow \pi^+\pi^-$ and intermediate $\Delta$-resonance
channels. Furthermore, motivated by recent $\pi\pi$ phase-shift
analyses which increasingly show evidence of s-wave contributions from
the $\sigma$ meson, the quasifree $\sigma$-decay channel
$$
vi)\ \ \ \gamma + {^3H}e \rightarrow \sigma ppn \rightarrow
\pi^+\pi^- \,{^3H}e
$$
has been added, with the $\sigma$ mass and width parameters from
Ref.~\cite{sigma}.

Last, $\rho^0 \rightarrow \pi^+\pi^-$ photoproduction has been
simulated by means of five distinct generators, namely
$$
\gamma + ^3He \rightarrow \left\{ \begin{array}{l}
{\left. \begin{array}{l}
{vii)\ \ \rho^0 + {^3H}e \rightarrow \pi^+\pi^- \,{^3H}e}\\
{viii)\ \rho^0ppn \rightarrow \pi^+\pi^-ppn}\\
{ix)\ \ \ \rho^0ppn \rightarrow \pi^+\pi^-ppn + (FSI)}
\end{array} \right\} \ \ \ \ \ m_{\rho^0}\,=\,770\, MeV/c^2}\\
{\left. \begin{array}{l}
{x)\ \ \ \ \rho^0 + {^3H}e \rightarrow \pi^+\pi^- \,{^3H}e}\\
{xi)\ \ \ \rho^0ppn \rightarrow \pi^+\pi^-ppn}
\end{array} \right\} \ \ 500\, MeV/c^2\, \le m^*_{\rho^0} \le \,725\, MeV/c^2}
\end{array} \right.
$$ where to channels $vii)$-$ix)$ and $x)-xi)$ are ascribed $\rho^0$
decay outside, and inside the nuclear medium, respectively, the latter
probing the medium effect on the $\rho^0$ mass. The breakup channels
have the reaction taking place on a single nucleon, subject to its
Fermi motion, with the remaining two nucleons as spectators.  Final
state $\pi N$ interaction (FSI) with one of the two spectator nucleons
is included in channel $ix)$.

\subsection{Fitting procedure}
\label{sec:fp}

In modelling $\pi^+\pi^-$ photoproduction on nuclei, the distributions
of five kinematical observables (Section~\ref{sec:obs}) were
simultaneously fitted to the data in Refs.~\cite{watts,lolos,hub}. These are
$$
\begin{array}{l}
{1.\ \ {\rm the\ dipion\ invariant\ mass\ }m_{\pi^+\pi^-}}\\
{2.\ \ {\rm the\ laboratory\ production\ angle\ of\ the\ dipion\ system\ }
\Theta_{IM}}\\ 
{3.\ \ {\rm the\ missing\ mass\ }m_{miss}}\\
{4.\ \ {\rm the\ missing\ momentum\ }p_{miss}}\\
{5.\ \ {\rm the\ }\pi^+ -\pi^- {\rm laboratory\ opening\ angle\ }
\vartheta_{\pi^+\pi^-}} 
\end{array}
$$
to which one additional kinematical observable has presently been added
(Section~\ref{sec:le1}), namely
$$
6.\ \ {\rm the\ } \pi^+ {\rm production\ angle\ in\ the\ dipion\ rest\ frame\ }
cos\theta^*_{\pi^+}\ .
$$

Moreover, in Ref.~\cite{watts} it was determined that dividing the
data sample in $\Delta E_\gamma$ = 80 MeV bins provided the optimal
compromise between the presumed constancy of the reaction matrix
elements, implicit in the MC simulations which only depend on the
kinematics, and the requirement of sufficient statistics. The $\Delta
E_\gamma$ partitioning of the data is necessary in order to account
for the varying energy dependence of the $\pi^+\pi^-$ cross sections
from each of the individual production mechanisms. The 80~MeV binning
in E$_\gamma$ has been retained, resulting in four sectors of the data
sample from 800 to 1120 MeV, to be referred by their respective
central E$_\gamma$ values (840, 920, 1000, and 1080 $\pm$40~MeV).

The addition of the cos$\theta^*_{\pi^+}$ distribution, though not
noticeably affecting the overall quality of the fit, nonetheless
provides an additional physical constraint which aids the MC fitting
algorithm to converge to a more realistic solution. In particular,
this kinematical observable uniquely captures a characteristic feature
of the contributing mechanisms, which may be classified into three
types according to their respective dependence on
cos$\theta^*_{\pi^+}$ (Fig.~\ref{figacc}): 

\begin{itemize} 
\item The channel of interest, diffractive $\rho^0\rightarrow
\pi^+\pi^-$, is unique in producing two p-wave correlated pions. The
spectrometer response to this mechanism is consistent with the $J=1$
dependence, and the deviation from the anticipated
cos$^2\theta^*_{\pi^+}$ distribution towards $\sim \pm$1 reflects the
acceptance cut, stemming from the kinematical conditions, set-up
geometry, and the two-pion detection requirement (compare the solid
curve of Fig.~\ref{figacc} with Figs.~\ref{figcos}b,c).
\item The background hadronic channels $i)-iv)$ of
Section~\ref{sec:pcs} produce two uncorrelated pions at two or more
reaction vertices. The angular correlations of these pions are
averaged out over 4$\pi$ sr in simulations, resulting in featureless
cos$\theta^*_{\pi^+}$ spectra. The spectrometer geometry, however,
suppresses the pion acceptance away from cos$\theta^*_{\pi^+}$=0 (see
e.g. the dashed curve of Fig.~\ref{figacc} for the single-$\Delta$
production channel). This is a consequence of the fact that channels
$i)-iv)$ involve the decay of intermediate baryonic resonances,
accompanied by energetic nucleons.
\item Three-pion production and the quasi-elastic $\sigma$ process,
$v)$ and $vi)$ of Section~\ref{sec:pcs}, are characterized by
featureless cos$\theta^*_{\pi^+}$ acceptances, as no energetic
nucleons are emitted (dotted curve of Fig.~\ref{figacc}).
\end{itemize}

The combination of improvements relative to the analysis of
Ref.~\cite{lolos}, namely, accounting for the angular momentum in the
$\Delta$ and $\rho^0$ channels, and imposing additional physical
constraints via the new kinematical observable cos$\theta^*_{\pi^+}$,
resulted in a more accurate treatment of the process-dependent
spectrometer acceptances.

The data have been subjected to two additional cuts, one of which
enhances the $\rho^0$ relative to the background channels, and the
other facilitates the separation of the bound ${^3H}e$ from the
breakup $ppn$ final states. The former is a $\pi^+\pi^-$ opening-angle
cut determined from MC simulations, namely 70$^\circ \le
\vartheta_{\pi^+\pi^-} \le$ 180$^\circ$, eliminating two-pion events
that could not have been produced back-to-back from the $\rho^0$ decay
(Fig.~\ref{figcuts}a). This cut is most effective at the higher end of
photon energies covered by the experiment, where the $\rho^0$
identification becomes difficult by an increasingly deteriorating
$\rho^0$-to-background ratio with increasing E$_\gamma$. The latter
cut (Fig.~\ref{figcuts}b) separates events with missing mass in the
proximity of the target mass $m_{{^3H}e}\,\approx\,2.8\ GeV/c^2$
(i.e. $2700\ MeV/c^2 < m_{miss} < 2865\ MeV/c^2$), from those
associated with the breakup of the target nucleus to $ppn$ in the
final state ($2865\ MeV/c^2 < m_{miss} < 3050\ MeV/c^2$). The
combination of the two types of cuts is applied to three of the
kinematical observables, resulting in six additional spectra, besides
the unselected $\pi^+\pi^-$ distributions 1-6 enumerated
earlier. These are:
$$
\begin{array}{l}
{\left. \begin{array}{l}
{7.\ \ m_{\pi^+\pi^-}}\\
{8.\ \ p_{miss}}\\
{9.\ \ cos\theta^*_{\pi^+}}
\end{array} \right\} \ \ \rho^0- {\rm enhanced\ low\ missing-mass\ data}}\\
{ }\\
{\left. \begin{array}{l}
{10.\ m_{\pi^+\pi^-}}\\
{11.\ p_{miss}}\\
{12.\ cos\theta^*_{\pi^+}}
\end{array} \right\} \ \ \rho^0- {\rm enhanced\ high\ missing-mass\ data}}
\end{array}
$$

While the contributing $\rho^0$ photoproduction mechanisms may not be
experimentally distinguishable (Section~\ref{sec:intro}), the cuts aim
at partitioning the data in biased samples favoring processes which
are more prone to probing either the nuclear mean-field effect
(i.e. coherent and exclusive quasifree photoproduction via the
distributions 7-9), or a possible nucleonic effect (i.e. the breakup
quasifree channel via the spectra 10-12). The three kinematical
observables which were subjected to the cuts, namely, the dipion
invariant mass, missing momentum, and dipion-cm $\pi^+$
production-angle, were selected empirically from the kinematical
observables 1-6 as more sensitive to the $\rho^0$ mass, and therefore
more susceptible to possible medium effects.

With a range of m$^*_{\rho^0}$ values, traversing the region 500-725
MeV/c$^2$, and each value kept common for the unselected, as well as
the ${^3H}e$ and $ppn$ selected data, the twelve simulated spectra
were fitted simultaneously to the corresponding experimental ones, by
minimizing a standard $\chi^2$ function with the strengths of the
eleven individual processes $i)-xi)$ (Section~\ref{sec:pcs}) as the
fit parameters. The four $\Delta$E$_\gamma = 80$~MeV bins were fitted
independently. The optimal m$^*_{\rho^0}$ for each bin is that
corresponding to the minimum value of the $\chi^2$ function
(Section~\ref{sec:mmm}).

In summary, the MC fitting procedure satisfies the following requirements:
\begin{itemize}
\item The $\Delta$E$_\gamma$ = 80 MeV binning restricts the energy
dependence of the participating processes to the narrowest bin
possible without loss of sufficient statistics.
\item The six kinematical observables utilized in the fitting are
complementary, and account for different physical attributes of the
data sample. This imposes far more stringent constraints than an
analysis based on only the invariant mass distribution, as in the case
of the CERES data \cite{ceres}.
\item The simultaneous fitting of selected and unselected data aims at
isolating a strong signal from data samples most responsive to
possible $\rho^0$ mass modifications, while also incorporating in the
fit the bulk of $\pi^+\pi^-$ events produced in processes less
sensitive to such effects.
\end{itemize}

The outcome of this procedure is the medium-modified $\rho^0$ meson mass.

\section{RESULTS}
\label{sec:data}

\subsection{Quality of fit and uncertainties}
\label{sec:mmm}

Beyond the statistical and other experimental uncertainties which are
folded into the calibration and analysis of the measured $\pi^+\pi^-$
yields, additional sources of uncertainty are generated by the MC
simulations and fitting algorithm. i) The MC event generators depend
exclusively on kinematical parameters, neglecting other aspects of the
interaction. As an example, the channel $\gamma + {^3}He \rightarrow
(np)_{sp}\Delta\pi$ is modelled as $\gamma +{^3}He \rightarrow
(np)_{sp} \Delta^{++}\pi^-$, normalized by an isospin scaling factor
to account for the remaining hadronic charge states. This introduces
an uncertainty in the amplitude of this process, similar to which are
also present in other background hadronic channels. ii) Independent
total cross-section measurements of the constituent reactions, serving
to anchor their relative strengths in the full $\pi^+\pi^-$ production
process, are sparse (e.g. the quasi-elastic $\sigma$ channel). The
strength for some of the individual channels was inferred from
indirect sources. iii) Additional quasifree channels and independent
measurements to fix their strength are required in order to extract
precision cross sections for the background hadronic channels from the
data, processes which merit attention on their own behalf. This may
become possible following the analysis of three charged-particle
events from TAGX experiments, currently in progress.

Despite the caveats above, the medium-modified $\rho^0$ masses
extracted from the MC fitting procedure have remained remarkably
stable with respect to variations of the strength parameters of the
constituent reactions, within each of the four $\Delta E_\gamma$
bins. This is all the more significant considering that the data are
fit simultaneously for twelve spectra of six kinematical
observables. In conjunction with the direct $J=1$ fingerprint
discussed earlier, the insensitivity of the fit, within physical
constraints, adds confidence to the premise that the medium-modified
$\rho^0$-meson masses extracted from the MC simulations reflect
genuine features of the data sample.

Following the procedure discussed in Section~\ref{sec:mc}, several MC
fits have been performed with m$^*_{\rho^0}$ taking on values from a
mesh in the range 500-725 MeV/c$^2$ (Fig.~\ref{figch2}). The steepness
of the $\chi^2$-vs-m$^*_{\rho^0}$ curve is indicative of the
sensitivity of the data sample to $\rho^0$ mass modifications, within
each of the four $\Delta E_\gamma$ bins. Using this as a qualitative
criterion, the fitting of the 840 and 920 MeV samples, and to a lesser
extent of the 1000 MeV sample, have converged to a ``best''
m$^*_{\rho^0}$, whereas the fit for the 1080 MeV bin is essentially
insensitive to variations of m$^*_{\rho^0}$ (Fig.~\ref{figch2}).

For each of the 840, 920, and 1000 MeV bins individually, the MC fits
(dark circles, triangles and squares respectively in
Fig.~\ref{figch2}) yield the best m$^*_{\rho^0}$ and an estimate of
its uncertainty. This is achieved as follows: a) The uncertainty
$\sigma_{\chi^2}$ is assumed common within each bin. This is justified
by the fact that both the data set and the reaction matrix elements
used by the MC algorithm are common within each bin. b) A polynomial
expansion of the MC $\chi^2$ about (m$^*_{\rho^0}$-m$_0$), with m$_0$
in the proximity of the apparent minimum, is assumed to describe well
the dependence of the former on the latter. Subsequently a new
${\chi'}^2$ minimization yields the rank and the coefficients of the
polynomial. A third order polynomial gives the best result for the two
lower-energy bins, and it is also assumed to provide the best
description of the parent population for the 1000 MeV data
sample. This procedure also leads to conservative estimates for
$\sigma_{\chi^2}$, in particular 0.019, 0.013 and 0.017 for the 840,
920, and 1000 MeV bins, respectively. c) The polynomial coefficients
and $\sigma_{\chi^2}$ estimates are used to derive the optimal
m$^*_{\rho^0}$ and an estimate of its uncertainty for each bin. These
are:
\begin{itemize}
\item m$^*_{\rho^0}$ = 642$\pm$40 MeV/c$^2$, 800 MeV$\le$E$_\gamma$$\le$880 MeV
\item m$^*_{\rho^0}$ = 669$\pm$32 MeV/c$^2$, 880 MeV$\le$E$_\gamma$$\le$960 MeV
\item m$^*_{\rho^0}$ = 682$\pm$56 MeV/c$^2$, 960 MeV$\le$E$_\gamma$$\le$1040 MeV
\end{itemize}

The improvement in the medium-modified over the unmodified $\rho^0$
mass fits is evidenced to varied degrees in all the fitted
spectra. The order of the improvement is illustrated for the full
(unselected) data in the dipion invariant-mass, $\pi^+$-$\pi^-$
laboratory opening-angle, and missing-momentum distributions for the
840$\pm$40 MeV and 920$\pm$40 MeV bins, which are most affected by
modifications of the $\rho^0$ mass (Figs.~\ref{fig840},\ref{fig920}).

The dipion invariant mass spectra are not the most sensitive to
variations of the $\rho^0$ mass, among the twelve distributions of six
kinematical observables that were employed in the fit. This is seen,
for example, in comparing the improvement between the unmodified- and
modified-mass dipion invariant-mass with the opening-angle and
missing-momentum spectra (Figs.~\ref{fig840},\ref{fig920}), where the
latter two observables are seen to display a greater sensitivity. This
underlines the advantage of a fitting procedure that utilizes multiple
complementary kinematical observables as opposed to only the invariant
mass, thus capturing additional attributes of the physical process,
and resulting in a more realistic analysis.

\subsection{Discussion}
\label{sec:int}

\begin{figure}
\epsfig{file=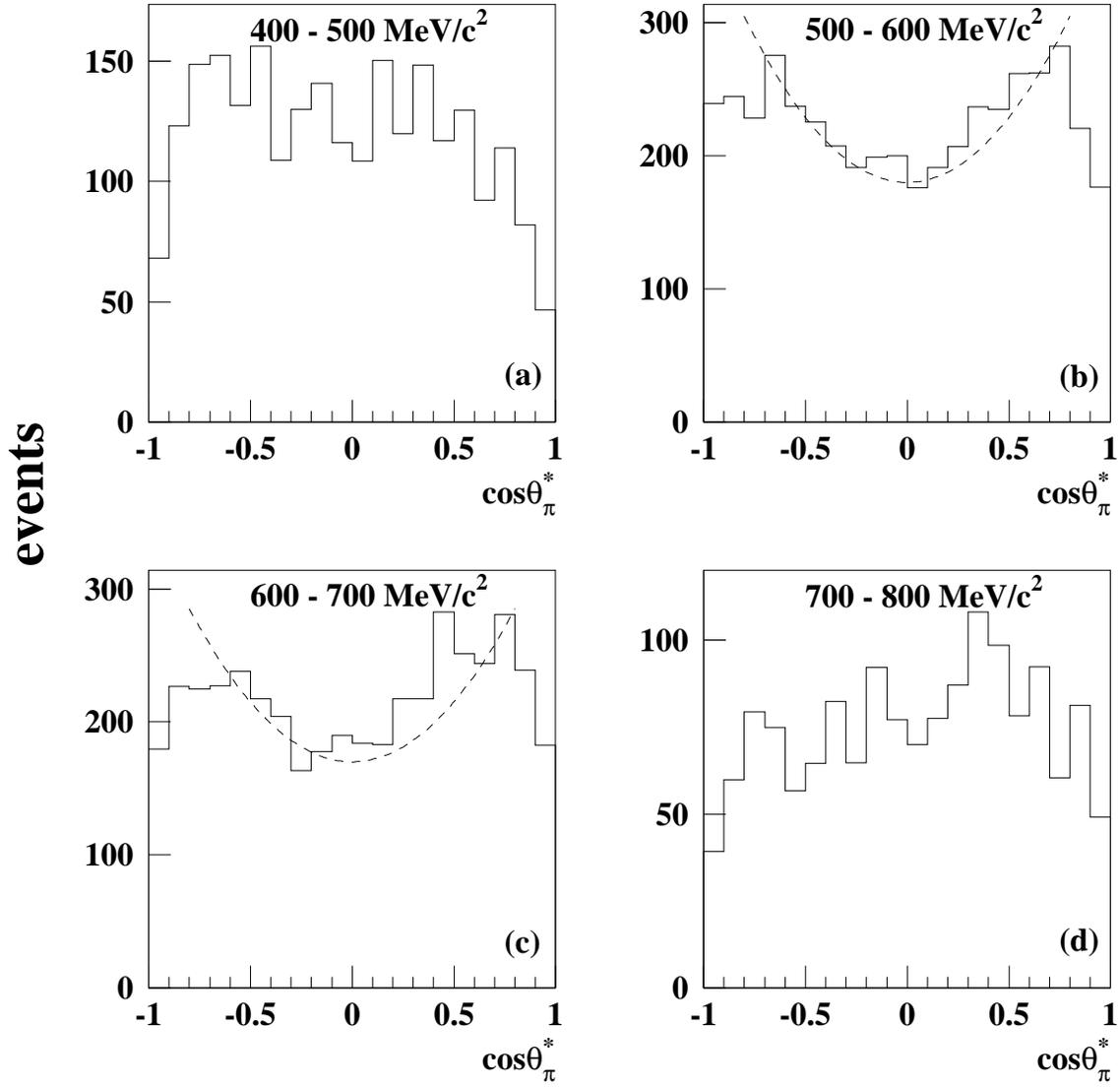}
\caption{The cos$\theta^*_{\pi^+}$ distribution captures the $J=1$
signature of the $\rho^0\rightarrow\pi^+\pi^-$ decay. Panels b) and c)
display the A+B\,cos$^2\theta^*_{\pi^+}$ dependence (dashed curve),
expected on the basis of the $J=1$ angular momentum, and the deviation
towards $\pm$1 is due to the spectrometer acceptance.}
\label{figcos}
\end{figure}

\begin{figure}
\epsfig{file=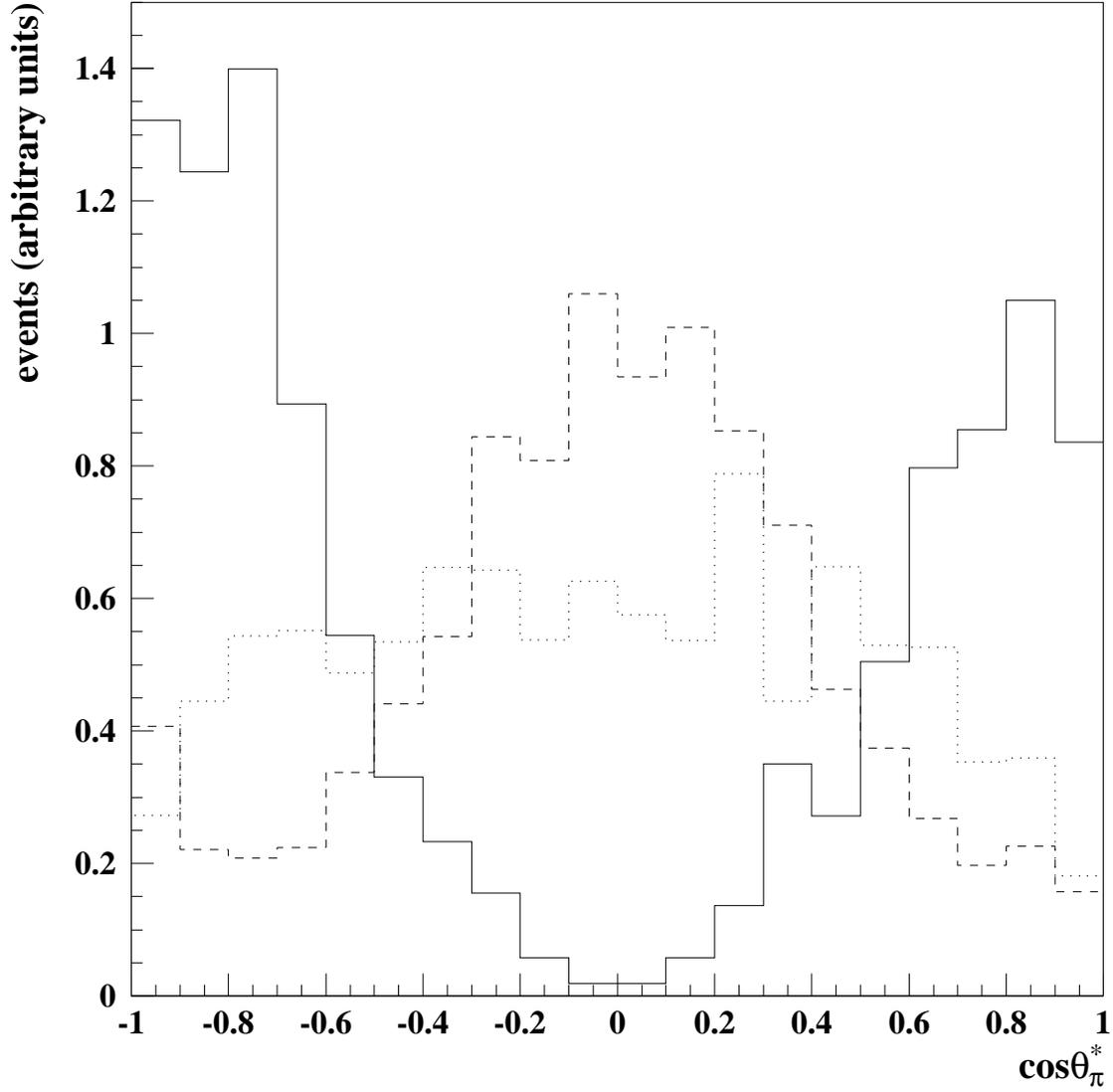}
\caption{Simulated events from different processes, taking into
account the spectrometer acceptance, have one of three characteristic
cos$\theta^*_{\pi^+}$ profiles. The simulations shown are differential
cross sections normalized to unity at E$_\gamma$=920 MeV, and they
display the cos$\theta^*_{\pi^+}$ dependence for three representative
reactions: a) $\gamma + {^3}He \rightarrow \rho^0 ppn \rightarrow
\pi^+\pi^-ppn$ with the nominal m$_{\rho^0}$ (solid curve), b) $\gamma
+ {^3}He \rightarrow \pi^-\Delta^{++}pn \rightarrow \pi^+\pi^-ppn$
(dashes) and c) $\gamma + {^3}He \rightarrow \sigma ppn \rightarrow
\pi^+\pi^-ppn$ (dots).}
\label{figacc}
\end{figure}

\begin{figure}
\epsfig{file=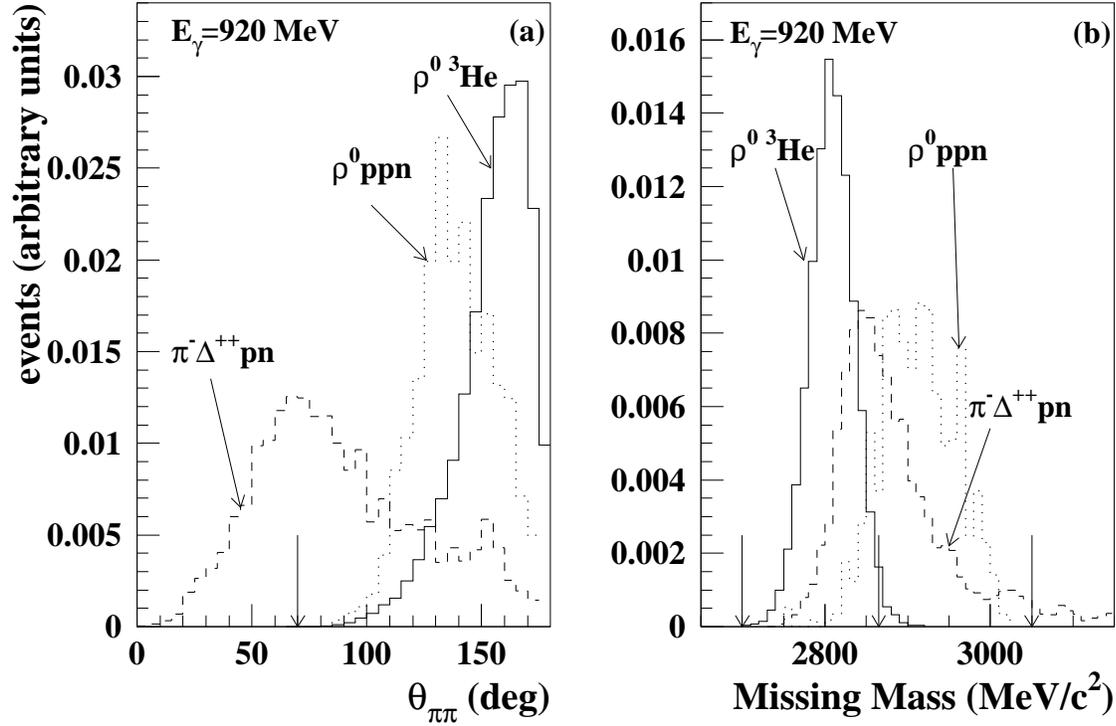}
\caption{The effects of the a) $\pi^+$-$\pi^-$ opening-angle and b)
missing-mass cuts, indicated as arrows in the figure, are shown for
the respective simulated spectra of the $\rho^0\,{^3H}e$, $\rho^0 ppn$
and $\pi^-\Delta^{++}pn$ channels (solid, dotted and dashed lines
respectively). The calculations are for E$_\gamma$=920 MeV and the
integral of each distribution is normalized to unity.}
\label{figcuts}
\end{figure}

\begin{figure}
\epsfig{file=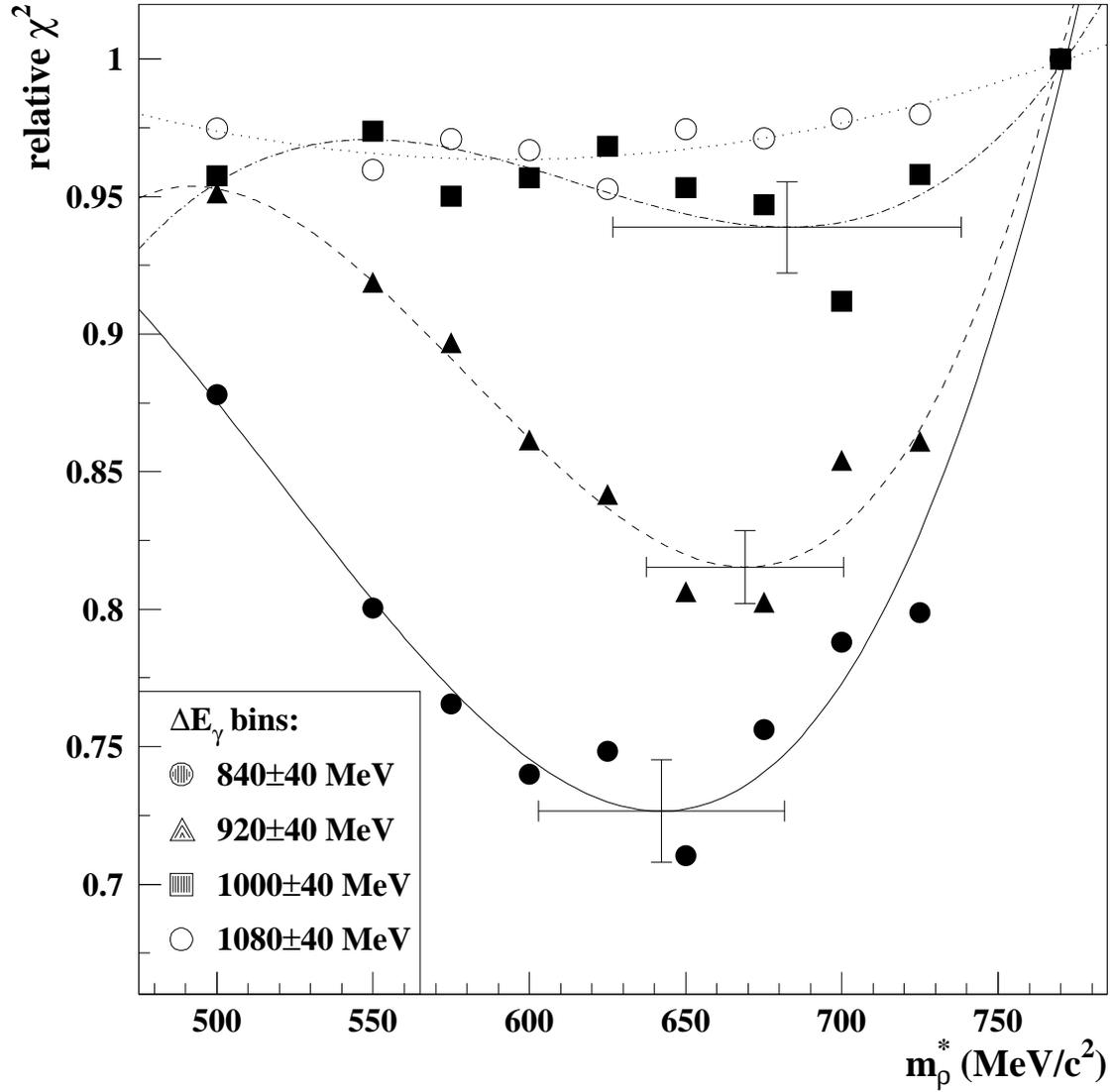}
\caption{The dependence of the $\chi^2$ function of the MC fits on the
variation of the $\rho^0$ mass is shown for the four $\Delta
E_\gamma$=80 MeV bins. The MC calculations are indicated as points,
and fits to third-order polynomials as curves. The fitting procedure
yields the best m$^*_{\rho^0}$ per $\Delta E_\gamma$ bin, and
estimates of the respective uncertainties.}
\label{figch2}
\end{figure}

\begin{figure}
\epsfig{file=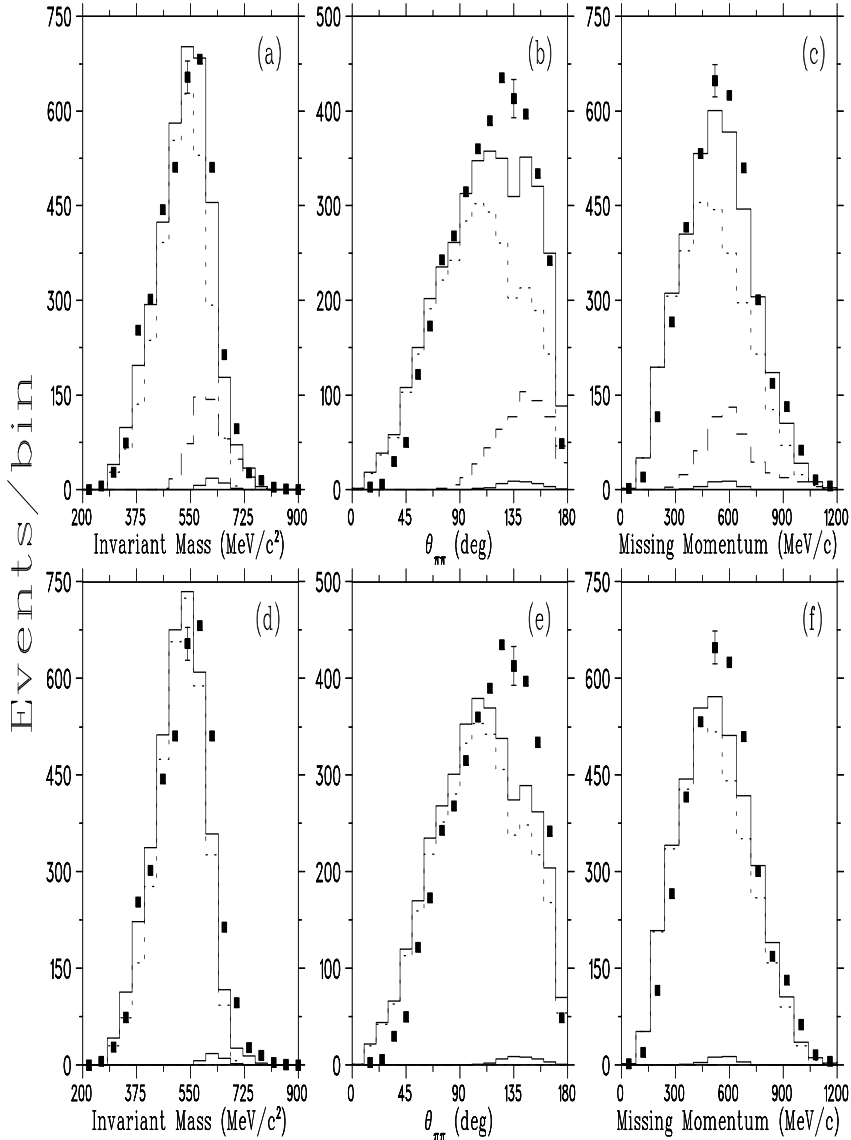,width=4.4in,height=6.5in,angle=90}
\caption{The dipion invariant-mass (panels a, d), $\pi^+$-$\pi^-$
laboratory opening-angle (panels b, e), and missing-momentum (panels
c, f) spectra are shown for the full (unselected) data at
E$_\gamma$=840 MeV. The solid squares are the data, showing only one
representative error bar for clarity.  The remaining error bars are
comparable and have been included in the fitting. The MC fits shown
are: full reaction (upper solid lines), non-$\rho^0$ background
processes (dotted lines), $\rho^0$ contributions with unshifted mass
(lower solid lines), and $\rho^0$ contributions with
m$^*_{\rho^0}$=650 MeV/c$^2$ (dashed lines). The latter are included
only in the calculations represented by the top panels (a, b, c).  The
improvement from bottom to top reflects the effect of including the
$\rho^0$ medium modifications.  This improvement is more pronounced
for the $\rho^0$ selected data set, not shown.}
\label{fig840}
\end{figure}

\begin{figure}
\epsfig{file=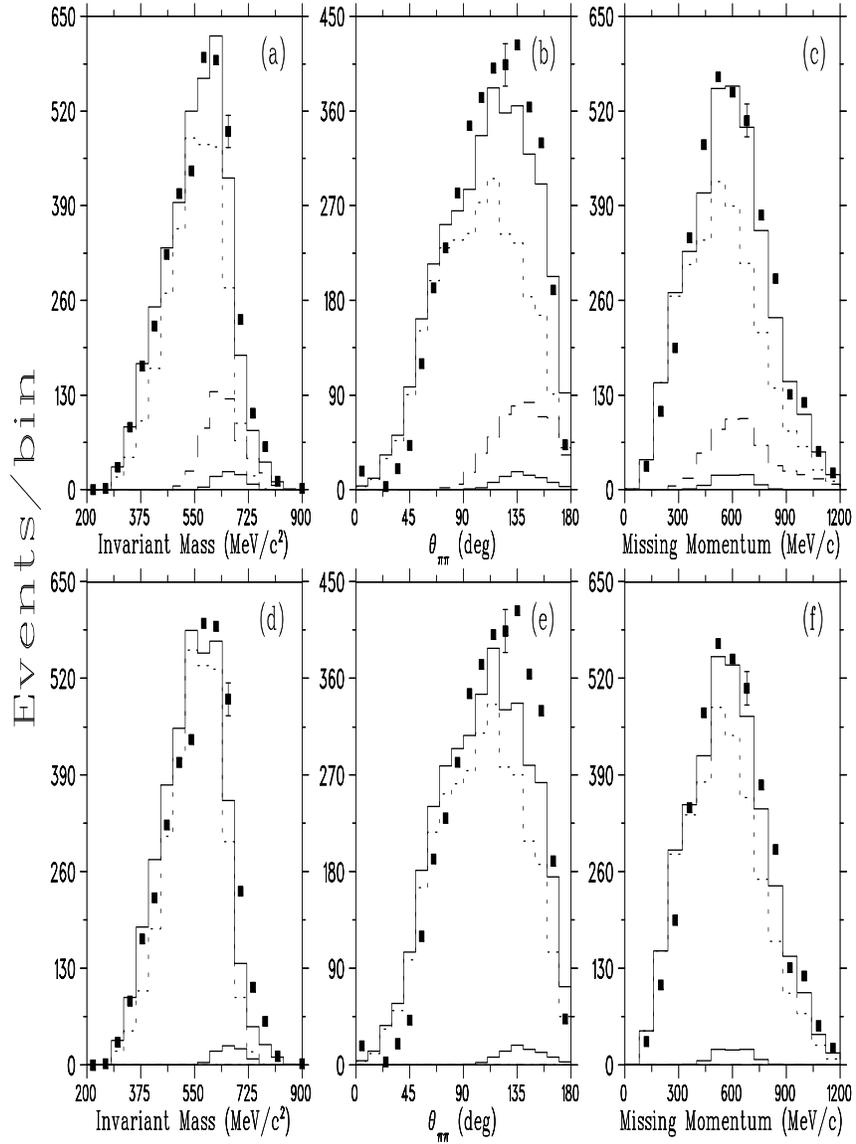,width=4.4in,height=6.5in,angle=90}
\caption{The data and curves shown are as in
Fig.~\protect{\ref{fig840}}, but for the E$_\gamma$=920 MeV bin, and
m$^*_{\rho^0}$=675 MeV/c$^2$. It is stressed that the fitting was done
independently for each of the four $\Delta$E$_\gamma$=80 MeV/c$^2$
bins, as discussed in the text.}
\label{fig920}
\end{figure}

\begin{figure}
\epsfig{file=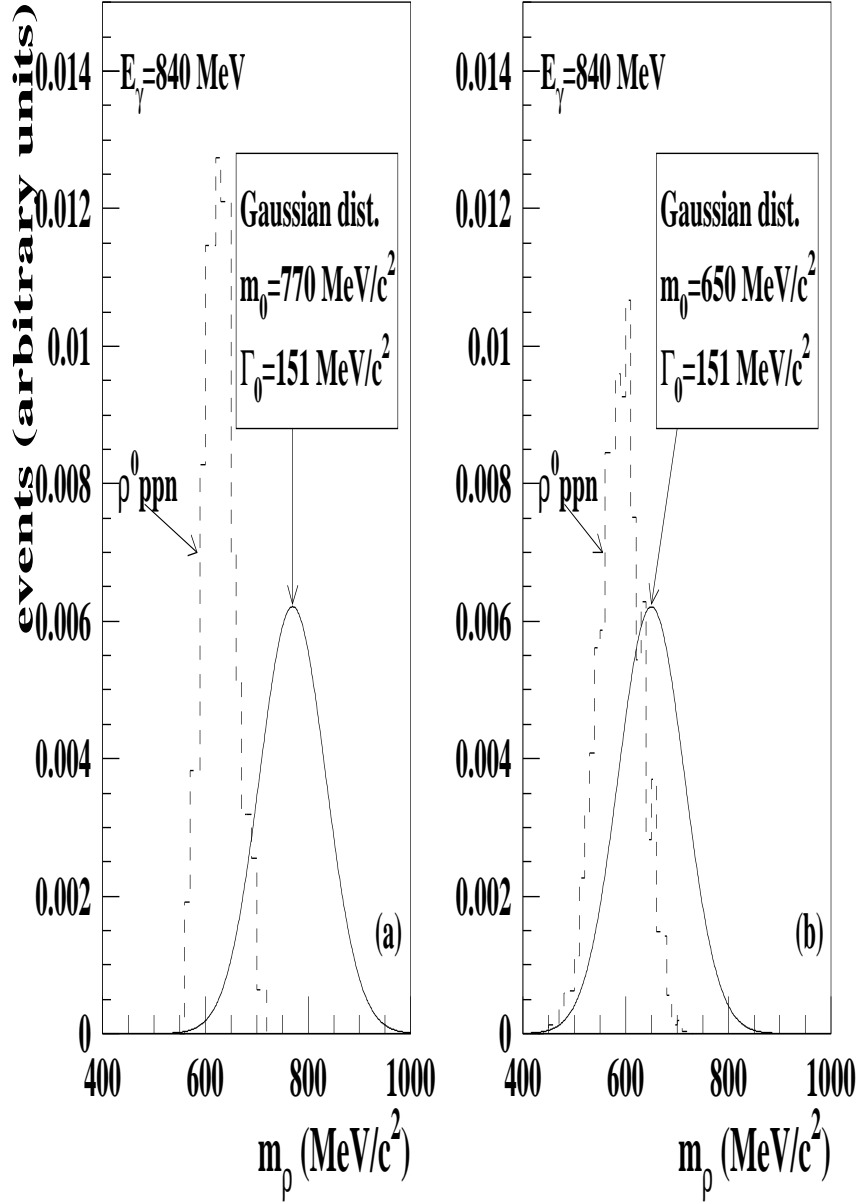}
\caption{The trivial $\rho^0$ mass shift due to kinematic phasespace
limitations is illustrated for the quasifree $\gamma + {^3H}e
\rightarrow \rho^0 ppn \rightarrow \pi^+\pi^-ppn$ process at
E$_\gamma$=840 MeV, a) for the nominal and b) for a lowered $\rho^0$
mass. The MC generator randomly selects $\rho^0$ masses from the
Gaussian distributions (solid curves). The histograms (dashed curves)
are the resulting m$_{\rho^0}$ spectra whose integrals are normalized to
unity, after verification that the reaction is kinematically feasible
with the selected mass and that the photoproduced pions are accepted
by the spectrometer.}
\label{figims}
\end{figure}

The $J$=1 angular momentum signal, discussed in Section~\ref{sec:le1}
(Fig~\ref{figcos}), as well as the dipion invariant mass spectra from
the fit (e.g. Figs.~\ref{fig840}a,d for E$_\gamma$=840 MeV), would
appear to indicate medium-modified $\rho^0$ masses which are actually
{\it lower} than the values quoted earlier (e.g. compare
m$^*_{\rho^0}$=642$\pm$40 MeV/c$^2$ for the 840 MeV bin with the
centroid suggested by the dashed curve of Fig.~\ref{fig840}a). This
apparent discrepancy is misleading, and has its origin in a trivial
effective mass lowering driven by phase-space, most pronounced at
lower photon energies. This is due to the fact that, at low photon
energies, only the lower wing of the $\rho^0$ mass distribution is
kinematically accessible (e.g. Fig.~\ref{figims} for E$_\gamma$=840
MeV and the $\rho^0ppn$ final state). The shape and the centroid of
the mass distribution for the resulting $\rho^0$ mesons is primarily
dictated by the kinematics, and to a lesser extent by the spectrometer
acceptance for the particular $\rho^0$-producing process. This effect
is implicit in the MC generated $\rho^0$ spectra, and it is by no
means sufficient to account for a mass lowering of the order
explicitly observed in the experiment. This is manifest, for example,
in comparing the lower solid curves of Figs.~\ref{fig840}a,d - for
which $\rho^0$ production is dominated by the $ppn$ final state - with
the $J$=1 fingerprint of the $\rho^0$ in Fig.~\ref{figcos}. The loci
of the former curves, consistent with the mass distribution indicated
by the histogram of Fig.~\ref{figims}a for E$_\gamma$=840 MeV and the
nominal $\rho^0$ mass, are far too high to be compatible with
Fig.~\ref{figcos}, with a $\rho^0$ signal peaking in the range of
500-600 MeV/c$^2$. In contrast, the histogram of Fig.~\ref{figims}b
for a lowered $\rho^0$ mass, and the dashed curve in
Fig~\ref{fig840}a, are consistent.

Alternate fits were also attempted, to possibly discern additional
features from the 840 and 920 MeV bin samples. Decoupling the
exclusive ${^3}He$ from the breakup $ppn$ final states with respect to
m$^*_{\rho^0}$ yielded identical masses for the 920~MeV bin, and a
somewhat smaller mass for the latter relative to the former channel
for the 840~MeV bin. The resulting improvement in the quality of fit,
however, is within the uncertainty estimate of $\sigma_{\chi^2}$. Fits
were also performed with $\Gamma^*_{\rho^0}$ fixed to the predicted
width at half nuclear density extracted for ${^3}He$ from
Ref.~\cite{weise}, about double the free width $\Gamma_{\rho^0}$. A
sizeable improvement in $\chi^2$ was noted with the modified width for
any mass, compared with the free mass and width case. In all cases,
however, the $\chi^2$ is 5-10\% larger with $\Gamma^*_{\rho^0}$ than
with $\Gamma_{\rho^0}$.  Moreover, the preferred m$^*_{\rho^0}$ is no
higher with the modified than with the free width. In summary, these
exploratory fits verify the preference in a reduced $\rho^0$ mass, but
are inconclusive, within the sensitivity of the data, as to whether a
width modification is {\it in addition} supported.

\begin{figure}
\epsfig{file=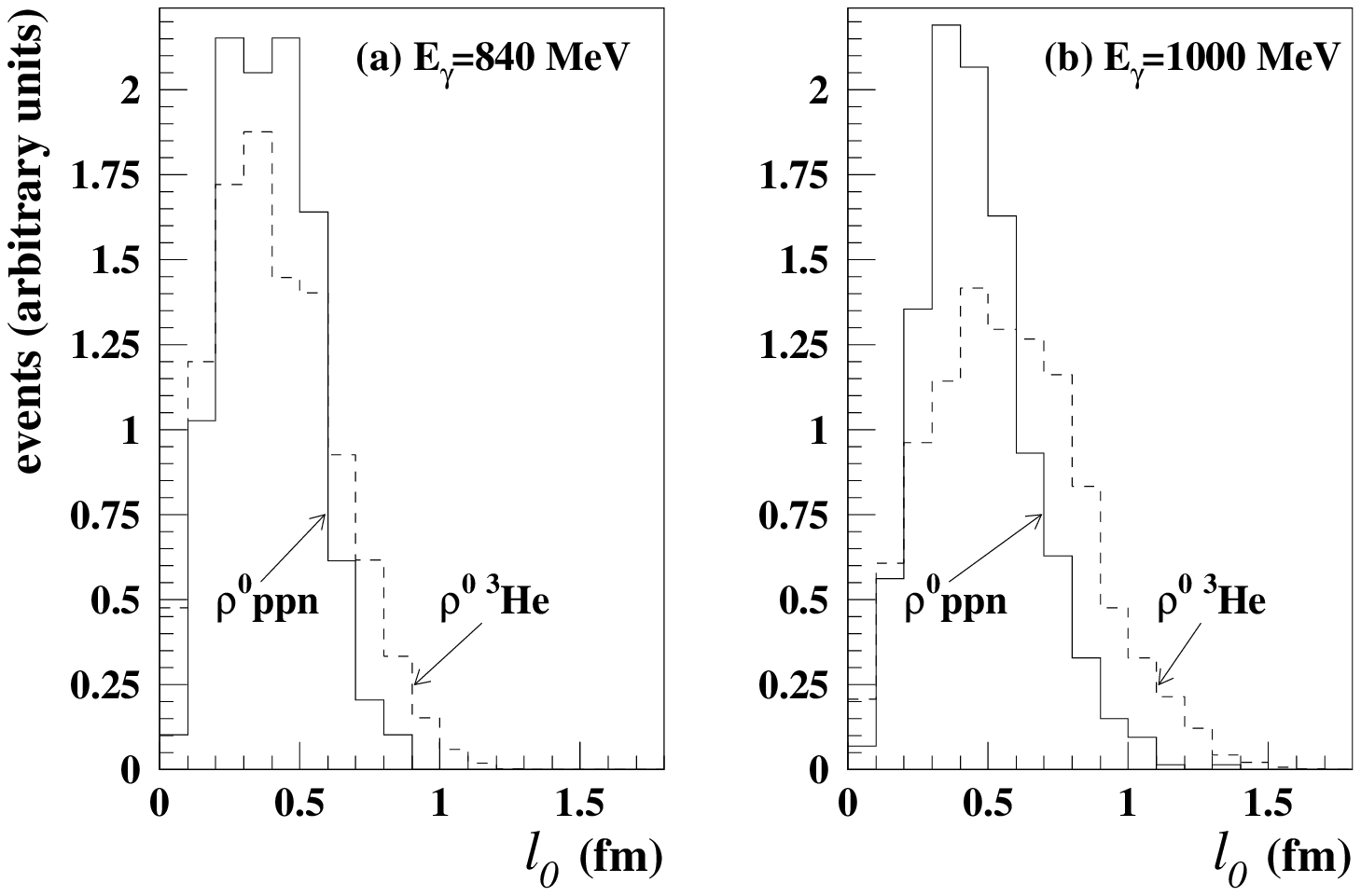}
\caption{The mean decay length $l_0$ distribution, of the $\rho^0 \rightarrow
\pi^+\pi^-$ decay, is indicative of the distances probed for vector-meson
modifications in this experiment. The $l_0$ spectra are illustrated
for a) E$_\gamma$=840 MeV and b) E$_\gamma$=1000 MeV. The histograms
are for MC generated events accepted by the spectrometer in the
$\gamma + {^3H}e \rightarrow \rho^0 p(pn)_{sp} \rightarrow
\pi^+\pi^-p(pn)_{sp}$ (solid curves) and $\gamma + {^3H}e \rightarrow
\rho^0\,{^3H}e\rightarrow \pi^+\pi^-\,{^3H}e$ (dashed curves)
reactions, with their integrals normalized to unity. The mean decay
length is for the former reaction in the rest frame of the
participating proton (the two remaining nucleons being spectators),
and for the latter in the rest frame of the ${^3H}e$ nucleus.}
\label{figdl0}
\end{figure}

The absence of a conclusive $\rho^0$ mass-modification dependence from
the 1000 and 1080 MeV bins, in contrast to the strong signal from the
840 and 920 MeV bins, is telling as well. Whereas the former are more
prone to probing the exclusive channel, and therefore the nuclear
mean-field medium effect, the latter are deeper into the subthreshold
region, dominated by the breakup channel, and probe distances
shorter than the nucleonic radius. To illustrate the range of the
$\rho^0$ processes in different energy regions, we consider the mean
decay length $l_0$ of the $\rho^0$
\begin{eqnarray}
&&l_0 = \beta t_0={{p\hbar}\over{m_{\rho^0}\Gamma_{\rho^0}c}}\nonumber \\
&&\beta = {p\over{m_{\rho^0}\gamma c}}\nonumber \\
&&t_0 = \gamma \tau_0 = \gamma {\hbar\over\Gamma_{\rho^0}}\ ,
\label{dl0}
\end{eqnarray}
with the nominal mass $m_{\rho^0}$ and width $\Gamma_{\rho^0}$ for the
$\rho^0$ meson, in the rest frame of either the interacting nucleon,
or the ${^3H}e$ nucleus, for the breakup or bound final state
(Fig.~\ref{figdl0}).  At low photon energies (e.g. E$_\gamma$=840 MeV,
Fig.~\ref{figdl0}a) the $ppn$ channel dominates $\rho^0$
photoproduction. This is verified by the fact that the low
missing-mass selected data represent less than 10\% of the total
unselected events, whereas the high missing-mass selected data
contribute the great majority of the $\rho^0$ events and their
corresponding distributions are generally consistent with the full
(unselected) data. With the $\rho^0$ mesons produced on the nucleon,
and a mean decay length of under $0.5$ fm for a substantial portion of
them, there is a large overlap of the volume traversed before their
decay and the nucleonic volume. To the extent that the medium
modifications of the vector-meson properties depend on the density of
the surrounding nuclear matter, it is conceivable that the induced
medium effect be dominated by the large densities encountered in the
interior of the nucleon, and consequently be far more pronounced than
a medium effect induced by nuclear fringe densities. At higher photon
energies (e.g. E$_\gamma$=1 GeV, Fig.~\ref{figdl0}b), both the $ppn$
and ${^3H}e$ final states contribute and the mean decay length
increases. Whereas nucleonic densities become increasingly
inaccessible, a large overlap of the volume traversed by the $\rho^0$
before its decay with the ${^3H}e$ nucleus may still induce a weaker
nuclear medium effect. Moreover, the possibility of a medium-induced
increase in the width $\Gamma_{\rho^0}$ is in favor of shorter decay
lengths (Eq.~\ref{dl0}), and therefore further increases the
likelihood of accessing regions of large densities either in the
nucleon, or the nuclear core.  The present results may therefore be
suggestive of a moderate medium effect in the realm of the nuclear
mean field at near-threshold photon energies, probing normal nuclear
densities, turning to a drastic reduction of the $\rho^0$ mass in the
deep subthreshold region, where the $\rho^0$ decay may be induced in
the proximity of the nucleon. These implications, however, need to be
further investigated and verified by higher precision and large solid
angle experiments \cite{sp}.

\section{Conclusions}
\label{sec:concl}

In summary, the kinematics and final state of the
${^3}He$($\gamma$,$\pi^+\pi^-$)X reaction have been studied with the
TAGX spectrometer and a tagged photon beam in the subthreshold
$\rho^0$ photoproduction region. The bound ${^3H}$e and breakup $ppn$
components of the $\rho^0$ channel have been investigated, aiming at
the distinction between a nuclear and a possible nucleonic medium
effect on the $\rho^0$ mass.  The $\rho^0$ channel has been aided by
the inherent selectivity of the TAGX spectrometer to coplanar
$\pi^+\pi^-$ events, and by the choice of ${^3}He$ as a target with
minimal FSI effects without suppression of the $\rho^0$ amplitude.  In
any case, the size of target has little effect in the subthreshold
region. The $J=1$ fingerprint of the $\rho^0$ has been observed in the
dipion invariant mass region 500-700 MeV/c$^2$, pointing to a
substantial reduction beyond a trivial phase-space governed apparent
lowering. This has been verified by MC simulations, incorporating the
exclusive and breakup $\rho^0$ channels, the latter both with and
without FSI, as well as background hadronic channels and s-wave
correlations. The extracted $\rho^0$ medium-modified masses,
642$\pm$40 MeV/c$^2$, 669$\pm$32 MeV/c$^2$, and 682$\pm$56 MeV/c$^2$
for the 840, 920, and 1000 MeV data bins, suggest a strong medium
effect in the deep subthreshold region, requiring large densities that
are incompatible with a nucleus as light as ${^3H}e$, but that are
more consistent with the probing of the nucleonic volume. The pattern
of decreasing sensitivity with increasing photon energy, with the 1080
MeV bin showing no evidence of a $\rho^0$ mass-modification signal,
hints a moderate mean-field nuclear effect at near-threshold
energies. The simulations are inconclusive regarding a medium-modified
$\rho^0$ width.

Further analysis currently in progress for photoproduction on heavier
targets (${^4}He$,${^{12}}C$) and for background contributions
(e.g. the $\Delta\pi$ channel) from TAGX experiments, as well as
future experiments planned for TJNAF \cite{sp}, may more accurately
isolate the $\rho^0 \rightarrow \pi^+\pi^-$ channel from the
background processes, and shed light on the nature of the medium
modifications on light vector-meson properties at the interface of
hadronic and quark matter.

The authors wish to thank H. Okuno and the staff of INS-ES for their
considerable help with the experiment.  Furthermore, the authors
acknowledge the very insightful discussions with P. Guichon, M.
Ericson, A. Thomas, and A. Williams during the Workshop on Hadrons in
Dense Matter (Adelaide, March 10-28, 1997), as well as the discussions
with N. Isgur at Jefferson Lab.  This work has been partially
supported by grants from INS-ES, INFN/Lecce, NSERC, and UFF/GWU.

\section*{FIGURES}

\newpage

\begin{references}
\bibitem[*]{kag} Present address: GSI, KPI/Leptonen, Planckstrasse 1,
64291 Darmstadt, Germany\\
E-mail: kagarlis@gsi.de
\bibitem[\dag]{garino} Present address: Department of Physics, University of
Evansville, Evansville, Indiana 47722
\bibitem[\ddag]{ins} Present name: KEK-Tanashi Division
\bibitem{casher} A. Casher, Phys. Lett. B {\bf 83}, 395 (1979).
\bibitem{pisar} R. Pisarsky, Phys. Lett B {\bf 110}, 155 (1982).
\bibitem{hat} T. Hatsuda and S. H. Lee, Phys. Rev. C {\bf 46}, R34 (1992).
\bibitem{geb1} G. E. Brown, Nucl. Phys. A{\bf 488}, 689 (1988)
\bibitem{geb2} G. E. Brown and C. M. Rho, Phys. Rev. Lett. {\bf 66},
  2720 (1991).
\bibitem{ceres} CERES Collaboration, G. Agakichiev {\it et al.}, Phys. Rev.
Lett. {\bf 75}, 1272 (1995);\\
 G. Agakichiev {\it et al.}, Phys. Lett. B, in print.
\bibitem{he3} HELIOS-3 Collaboration, M. Masera {\it et al.}, Nucl. Phys.
{\bf A590}, 93c (1995);\\
NA50 Collaboration, E. Scomparin {\it et al.}, {\it ibid.} {\bf A610},
331c (1996).
\bibitem{geb3} G. Q. Li, C. M. Ko and G. E. Brown, Phys. Rev. Lett. {\bf 75},
4007 (1995).
\bibitem{cas} W. Cassing, W. Ehehalt and C. M. Ko, Phys. Lett. {\bf B363},
35 (1995);\\
C. P. de los Heros, Ph.D. Dissertation, Weizmann Institute, 1996.
\bibitem{chanfray} R. Rapp, G. Chanfray and J. Wambach, Nucl. Phys.
{\bf A617}, 472 (1997).
\bibitem{ccf} W. Cassing, E. L. Bradkovskaya, R. Rapp, and J. Wambach,
Phys. Rev. C {\bf 57}, 916 (1998).
\bibitem{friman} B. Friman and H. J. Pirner, Nucl. Phys. {\bf A617},
496 (1997).
\bibitem{weise} F. Klingl, N. Kaiser, and W. Weise, Nucl. Phys. {\bf A624},
527 (1997).
\bibitem{rwbr} G. E. Brown, G. Q. Li, R. Rapp, M. Rho, and J. Wambach,
nucl-th/9806026, to be published in Acta Phys. Polon.
\bibitem{jpsi} NA50 Collaboration, M. Gonin {\it et al.},
Nucl. Phys. {\bf A610}, 404c (1996).
\bibitem{kaons} G. E. Brown {\it et al.}, Phys. Rev. Lett. {\bf 60},
2723 (1998), and references therein.
\bibitem{si} E. J. Stephenson {\it et al.}, Phys. Rev. Lett. {\bf 78},
1636 (1997).
\bibitem{isg} R. Koniuk and N. Isgur, Phys. Rev. D {\bf 21}, 1868 (1980).
\bibitem{sai} K. Saito, K. Tsushima and A. W. Thomas, Phys. Rev. C {\bf 56},
566 (1997).
\bibitem{maru} K. Maruyama {\it et al.}, Nucl. Instr. Meth. {\bf A376},
335 (1996).
\bibitem{watts} D. G. Watts {\it et al.}, Phys. Rev. C {\bf 55}, 1832 (1997).
\bibitem{crho} H. Alvensleben {\it et al.}, Nucl. Phys. {\bf B18}, 333 (1970).
\bibitem{lolos} G. J. Lolos {\it et al.}, Phys. Rev. Lett. {\bf 80},
241 (1998).
\bibitem{zisis} Z. Papandreou {\it et al.}, submitted to Phys. Rev. C.
\bibitem{hub} G. M. Huber, G. J. Lolos and Z. Papandreou, Phys. Rev. Lett.
{\bf 80}, 5285 (1998).
\bibitem{geb4} G. E. Brown, M. Buballa and M. Rho, Nucl. Phys. {\bf A609},
519 (1996).
\bibitem{guic} P. A. M. Guichon and M. Ericson, private communication.
\bibitem{sp} G.J. Lolos {\em et al.}, ``Medium Modifications of Vector 
Mesons in the Subthreshold Region'', JLab Proposal (submitted to PAC-15).
\bibitem{gar} G. Garino {\it et al.},  Nucl. Instr. Meth. {\bf
A388}, 100 (1997).
\bibitem{target} M. Harada {\it et al.}, Nucl. Instrum. Methods
Phys. Res. A {\bf 276}, 451 (1989);\\
S. Kato {\it et al.}, {\it ibid.} {\bf 290}, 315 (1990);\\
S. Kato {\it et al.}, {\it ibid.} {\bf 307}, 213 (1991).
\bibitem{iur} M. Iurescu, M.Sc. Thesis, University of Regina, 1997,
unpublished;\\ A. Weinerman, M.Sc. Thesis, University of Regina, 1996,
unpublished.
\bibitem{aki} A. Shinozaki, University of Regina TAGX Collaboration
Report, 1998, unpublished.
\bibitem{rhob} I. S. Hughes, {\it Elementary Particles}
(Cambridge University Press, Cambridge, 1985), 2nd ed., pp. 201-202.
\bibitem{rhoa} W. D. Walker, J. Carroll, A. Garfinkel, and B. Y. Oh,
Phys. Rev. Lett. {\bf 18}, 630 (1967).
\bibitem{saphir1} W. Scwille {\it et al.} (SAPHIR Collaboration), Nucl. Instr.
Meth. {\bf A344}, 470 (1994).
\bibitem{saphir2} F. J. Klein, Ph.D. Dissertation, Bonn IR-96-08 (1996).
\bibitem{mami1} A. Braghieri {\it et al.}, Phys. Lett. B {\bf 363}, 46 (1995).
\bibitem{mami2} F. H\"arter {\it et al.}, Phys. Lett. B {\bf 401}, 229 (1997).
\bibitem{ppp1} J. A. Gomez-Tejedor and E. Oset, Nucl. Phys. {\bf A571}, 667
(1994);\\ {\it ibid.} {\bf A600}, 413 (1997).
\bibitem{ppp2} L. Y. Murphy and J. M. Laget, DAPNIA/SPhN 95-42 (1995).
\bibitem{ppp3} K. Ochi, M. Hirata, and T. Takaki, Phys. Rev. C {\bf 56},
1472 (1997).
\bibitem{ssk} S. S. Kamalov and E. Oset, Nucl. Phys. {\bf A625}, 873 (1997).
\bibitem{bianchi} N. Bianchi {\it et al.}, Phys. Rev. C {\bf 54}, 1688 (1996).
\bibitem{dd1} M. Asai {\it et al.}, Z. Phys. A {\bf 344}, 335 (1993).
\bibitem{dd2} Y. Wada, for the SAPHIR Collaboration, Particles and Nuclei
International Conference, Williamsburg, Virginia, 1996 (unpublished).
\bibitem{on} J. A. Gomez-Tejedor, E. Oset, and H. Toki, Phys. Lett. B
{\bf 346}, 240 (1995).
\bibitem{saphir3} J. Hannappel, Ph.D. Dissertation, Bonn IR-96-04 (1996).
\bibitem{ppp} The ABBHHM Collaboration, Phys. Rev. {\bf 175}, 1669 (1968).
\bibitem{sod} D. Luke and P. Soding, in ``Symposium on Meson-, Photo-,
and Electroproduction at Low and Intermediate Energies,'' Vol. 59 of {\it
Springer Tracts in Modern Physics}, edited by G. H\"ohler (Springer-Verlag,
Berlin, 1971), p. 39.
\bibitem{sigma} R. M. Barnett {\it et al.}, Phys. Rev. D {\bf 54}, 1 (1996).
\end{references}
\end{document}